\documentclass[%
 reprint,
 amsmath,amssymb,
 aps,
]{revtex4-2}

\usepackage{braket}
\usepackage{graphicx}
\usepackage{dcolumn}
\usepackage{bm}

\begin{document}

\preprint{APS/123-QED}

\title{Tutorial:\\A practical guide to the alignment of defocused spatial light modulators for fast diffractive neural networks.}

\author{Guillaume N\oe{}tinger}
\thanks{These two authors contributed equally}
\author{Tim Tuuva}%
\thanks{These two authors contributed equally}
\author{Romain Fleury}%
 \email{romain.fleury@epfl.ch}
\affiliation{%
Laboratory of Wave Engineering, Department of Electrical Engineering, Ecole Polytechnique Fédérale de Lausanne (EPFL), 1015 Lausanne, Switzerland
}%

\date{\today}

\begin{abstract}
The conjugation of multiple spatial light modulators (SLMs) enables the construction of optical diffractive neural networks (DNNs). To accelerate training, which is limited by the low refresh rate of SLMs, spatial multiplexing of the input data across different spatial channels is possible, maximizing the number of available spatial degrees of freedom (DoFs). Precise alignment is required in order to ensure that the same physical operation is performed across each channel and thus the learning operation of the network. We present a semi-automatic procedure for this experimentally challenging alignment resulting in a pixel-level conjugation. It is scalable to any number of SLMs and may be useful in wavefront shaping setups where precise conjugation of SLMs is required, e.g. for the control of optical waves in phase and amplitude. The resulting setup functions as an optical DNN capable of processing hundreds of inputs simultaneously, thereby reducing training times and experimental noise through spatial averaging. We further present a characterization of the setup and an alignment method. 
\end{abstract}

\maketitle

\section{Introduction}


Advances in artificial intelligence (AI) have brought the field to the forefront of contemporary scientific research, with applications ranging from the generalization of large language models to the prediction of protein structures. To address more complex scientific or technical problems, neural networks are becoming increasingly complex \cite{HAI_Maalej24}. In the context of a widespread AI deployment, concerns are being raised about its ecological impact \cite{Dhar2020} and the predicted future limitations of computing capacity due to the saturation of Moore's law \cite{Mitchell16}. As a result, the previously overlooked domain of \emph{optical computing} \cite{AthalePsaltis16, Goodman87} is experiencing a revival with a rapidly growing community in recent years \cite{McMahon2023}.\\ 

\begin{figure*}[t!]
\centering
\includegraphics[width=0.9\textwidth]{"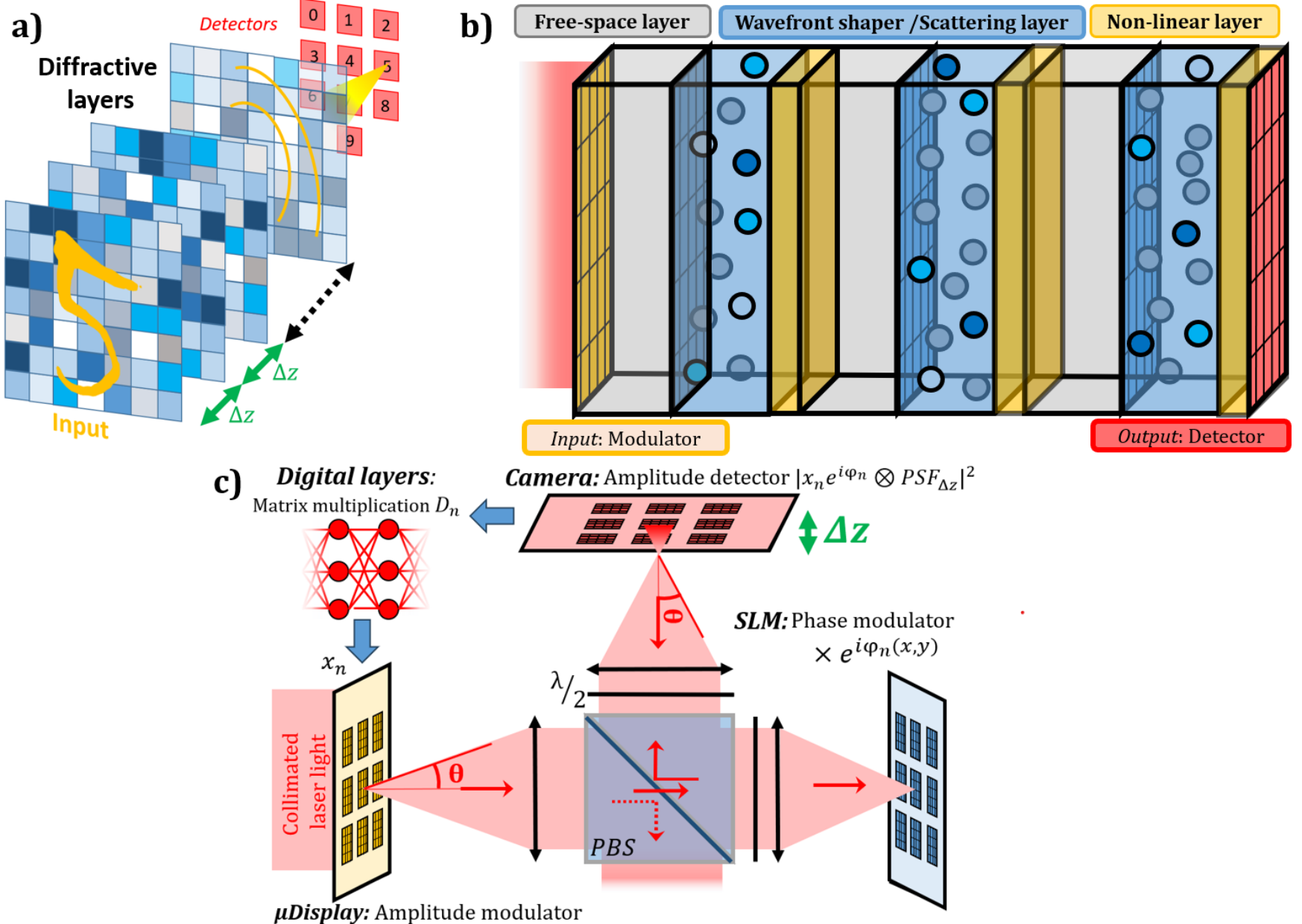"}
\caption{\label{fig:Ppe} \small{\textbf{An optical diffractive neural network} \textbf{(a)} The wave associated to the image of a given digit is sent on the corresponding detector by properly tuning the propagation media, forming a \textit{diffractive processor} [Ozcan] \textbf{(b)} A schematic example of a compact lithographed bulk photonic media able to perform inference passively in a compact volume. \textbf{(c)} Our setup mimicking this compact bulk photonic media. The proposed setup consist in conjugated amplitude and phase modulators sent to a defocused camera at a distance $\Delta z$ from the image plane, the resulting intensity is sent again onto the setup to be modulated again. Given the typical machine learning image size compared to the SLM resolution, multiple inputs can be displayed at the same time (here $3\times3$ images, $10 \times 10$ in our experiments) to speed up the acquisition. }}
\end{figure*}

An experiment based on the propagation of microwaves demonstrated that the propagation of a wave through a succession of carefully engineered layers can perform inference passively, similar to an artificial neural networks \cite{AllOpticalNN_LinOzcanScience2018}, as schematically depicted on Fig. \ref{fig:Ppe}.a. In this experiment, a microwave field is sent to a 3D-printed diffractive medium and, depending on the input pattern, is focused at a different position. This proves that waves can perform tasks currently reserved for numerical machine learning systems, such as \textit{classification}, but in a passive way with no energy cost associated to the input processing. These results sparked a novel research axis in so-called \emph{diffractive neural networks} (DNN) \cite{InferenceOpticsReview_Psaltis20, Zhou2021,Hu2024}. 
Using large wavelengths in the radio range allows to engineer precisely the propagation media relatively to the wavelength but results in cumbersome and slow experimental setups \cite{Zangeneh-Nejad2021}, with no available free-field source and no reconfigurability. Consequently, it becomes desirable to downscale the experiments to lightwaves, taking advantage of the favourable scaling of free-field optics \cite{McMahon2023}. The large number of reconfigurable spatial DoFs of SLMs or diffractive plates \cite{WangPiestun18} as well as the wide bandwidth of optical waves\cite{McMahon2023} and the impressive \textit{analog computing} abilities of full-field optics, such as the well-known 2D Fourier transform achieved at the speed of light by a simple lens \cite{FourierO}, are particularly interesting features for DNNs. This type of medium could be engineered at a miniature scale adapted to visible light wavelengths to provide fast inference at low energy cost \cite{Markovic2020}, as shown on Fig.\ref{fig:Ppe}.b. Assuming downsizing proportional to the wavelength reduction, i.e. from 750µm to 600nm, such a medium would occupy a volume of roughly 100µm$^3$. Theoretical and technical tools to engineer this type of optical medium are under developments. They involve many interconnected scientific questions, such as understanding wave propagation in complex media \cite{PrincipleScattering}, the engineering of optically reconfigurable media in the bulk could be achieved using inclusion of liquid crystals \cite{Nocentini2024} that are observable or individually addressable at reasonable speed using novel imaging techniques \cite{Najar2024}, and the addition of nonlinearities using specific materials \cite{AllOptNL_MiscuglioSorger18} or structure of the media \cite{PassiveDL_FeiGiganCaoArXiv}. Ultimately, the training of such devices \cite{NNtraining_Momeni24}, while taking into account the noise involved in any analog process, represents a significant challenge. In addition, the variability of manufacturing processes would induce some variability in the performance of different elements making the ability to train slightly different devices using the same algorithm crucial to this field. This concept is dubbed '\textit{mortal computing}' by Hinton \cite{FF_Hinton22} who proposed a solution in the form of a contrast learning algorithm called the \textit{Forward-Forward} algorithm which demonstrated its usefulness for training wave-based physical devices \cite{MomeniFF_Science2023}. In this article, we present the realization of a simple optical setup using two SLMs to mimick optical wave propagation in a reconfigurable media with non-linearities. It combines all these elements using off-the-shelf components and allows investigation of learning algorithms in a reasonable timeframe. We prove its operation using the hybrid training scheme of Forward-Forward. The procedure to align the two modulators as well as its characterization using intensity measurements only is carefully described.

An optical equivalent of this seminal microwave experiment involves a succession of SLMs \cite{Zhou2021, Chen2025} or the conjugation of successive SLMs with some \emph{defocus} simply obtained by translating the SLM in the image plane by a distance $\Delta z$ to account for the free-space propagation \cite{ChenGao23}, thereby mimicking diffraction in a reconfigurable layered medium \cite{Wanjura24, Kulce2021} except for the filtering of the angular spectrum from the limited numerical aperture (NA) of the lenses. Although this approach attracts theoretical interest \cite{savinson25} it has not, to our knowledge, been realized experimentally. The closest realization of this type of architecture is found in a multi-plane light converter (MPLC) configuration \cite{zhang2023MPLCpractical} which uses multiple reflections on the same SLM with a reflective layer \cite{JiaDing24, nPOLO_YildirmiPsaltisMoser24}. This is achieved at the expense of many DoFs as the SLM must be partitioned into several zones. There is also a coarser control of the diffraction between two reflections, since the spacing between the SLM and the mirror must be large enough for sending the input but small enough to ensure that diffraction remains moderate and does not direct too much optical power outside the collecting optics. Finally, there is a low tolerance to minor misalignments \cite{MPLC_Rocha25}. 

We assume the training is performed optically hence, the devices display and acquisition step is the limiting step of the training neglecting optical-to-electronic conversion and digital processing time. For a batch of 60 000 images, sent one at a time with a typical SLM refresh rate of 60Hz, such as the standard $28 \times 28$ pixels$^2$ MNIST image dataset used for classification and trained over 100 epochs, the training time would be roughly 27 hours for each \textit{layer} i.e. propagation through the optical setup. Similar orders of magnitude for processing times have been reported in the literature in systems with less DOFs \cite{OnChipPNNarbProgram_OnoderaMcMahon24}. A possibility to enhance the speed of this optical DNN is to use spatial multiplexing, by displaying many different inputs simultaneously making the most out of the high number of spatial DoFs. If it is possible to display 100 inputs at a time, the training time decreases to 27 minutes per layer for 100 epochs. This can be achieved by dividing each modulator surface into different regions of interest (ROIs). However, the use of multiplexing for a physical neural network raises significant concerns. It is necessary to ensure that each channel performs exactly the same physical operation as best as possible. For this, two corresponding RoIs of successive modulators need to be perfectly and similarly aligned. 
Indeed, this transition to optical waves is not straightforward: as the wavelength decreases to the visible range, e.g. approximately 500 nm, so does the typical tolerance for the mechanical precision and the level of control of the propagation medium. For a full-field setup, this presents significant experimental challenges in terms of alignment, as is the case for MPLCs \cite{MPLCLib_2025}.   

We propose an equivalent of a DNN created by aligning an amplitude and phase modulator, which is sent to a defocused camera (see Fig. \ref{fig:Ppe}.b).  The recorded data can then be sent through the setup again to mimic a multilayer neural network. The standard read-out non-linearity of the camera acts as a non-linearity added to each propagation through the setup, making this approach closer to a real digital neural network compared to a linear succession of diffractive plates. 

The conjugation is performed using manual alignments followed by an automated procedure for precise fine-tuning, resulting in pixel-wise alignment. This procedure is robust to mild defocus and is therefore suitable for diffractive neural networks as a single reconfigurable diffractive layer, with additional non-linearity arising from the intensity sensitivity of the camera, as in a convolutional neural network (CNN) (see Fig. \ref{fig:Ppe}.c). It is scalable to an arbitrary number of conjugated or defocused SLMs by applying the same procedure to each modulator. As such, it can be used, for instance, to better align the different diffractive zone of a MPLC to reduce the mismatch with numerical simulation. Furthermore, a defocused conjugation of multiple spatial light modulators, by enabling the tailoring of optical wavefronts over large fields of view in both the real domain and angular spectrum, may benefit other optical applications \cite{RoadmapWFS_Gigan22, RoadmapStructuredWaves23} such as the control of wave propagation in complex media \cite{MMFrothe24}, optical telecommunications \cite{dinc24Optoswitch}, holography \cite{Sheridan20} or imaging setups \cite{Maurer10, DynamicConfocal_Noetinger24}.

\section{The alignment protocol}

\subsection{The setup}

\begin{figure}[b!]
\centering
\includegraphics[width=1.05\columnwidth]{"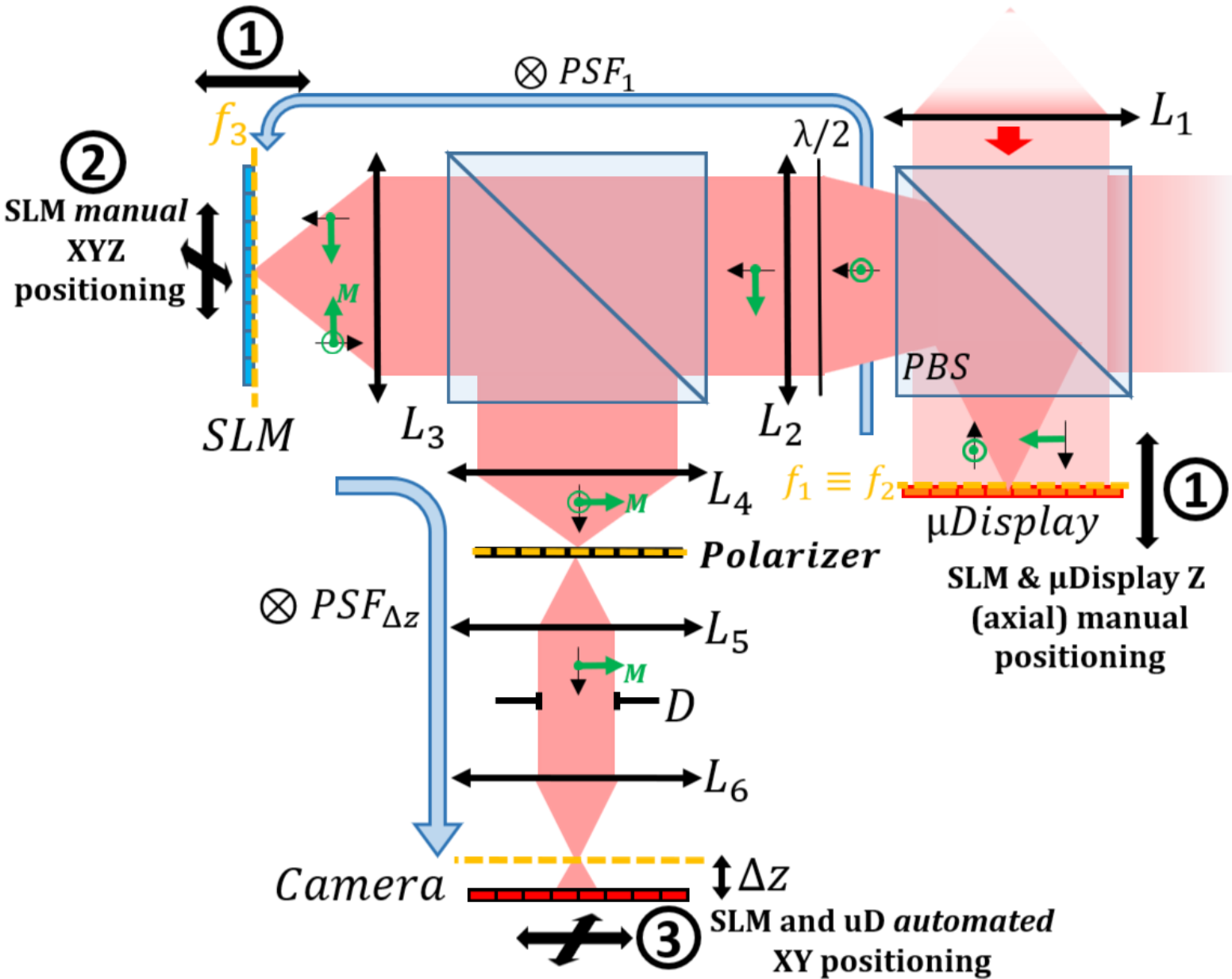"}
\caption{\label{fig:Setup} \textbf{The setup}. \footnotesize{ An amplitude modulator (the $\mu$Display) is conjugated to a phase modulator (the SLM) using a $4f$ system. The resulting field is sent to a defocused camera after filtering by a diaphragm $D$. The polarization is indicated by the green arrows. Only the horizontal polarization is modified by the SLM (green letter $M$). The remaining vertical polarization is filtered with a polarizer. The field on the SLM is the convolution of the µDisplay's image and a point spread function (PSF) $PSF_1$, the propagation to the defocused camera corresponds to a convolution by $PSF_{\Delta z}$. In practice, aberrations parasitic reflections and cross-talk have to be taken into account.}}
\end{figure}

The setup is schematically shown in Fig. \ref{fig:Ppe}.a and depicted in more detail in Fig. \ref{fig:Setup}. A red laser beam at 633 nm is expanded using a $4f$-system (lens $L_0$, not shown here, and $L_1$). The resulting plane wave is directed to an amplitude modulator (Holoeye microdisplay HEO 2220 CFS with a 4.5µm pixel pitch, hereafter referred to as the µDisplay) conjugated with a phase-only modulator (Holoeye ERIS with an 8µm pixel pitch, hereafter referred to as the SLM) using a \textit{4f}-system. The resulting beam is then imaged onto a camera after polarization filtering in a conjugate plane. A diaphragm allows filtering of high spatial frequencies in the Fourier plane to reduce the experimental noise. The width of this diaphragm does not affect the calibration algorithm as the PSF remains
centrosymmetric thus allowing to detect the centers of the ellipses whose images remain circular. Another use of this diaphragm may be to obtain a tunable speckle grain size (see SI). We define the point-spread functions (PSF) from the µDisplay to the SLM as $PSF_1$ and from the SLM to the camera as $PSF_{\Delta z}$.
A grid of  $10\times 10$ independent RoIs of size $64 \times 64$ pixels each is used. This corresponds to more than $4\cdot 10^5$ DoF. \\ 
To accelerate the training of the DNN, these RoIs are used as parallel channels, each performing exactly the same operation. This enables batch training for each capture, increasing the training speed by a factor equal to the number of RoIs, in our case $10^2$. This requires that the same wavefield from the µDisplay is sent to a diffractive layer of the SLM with the same phase law for each channel. 
With the parameters chosen here, assuming a 60 Hz display and acquisition frequency - which is standard for the selected modulators and camera - and an output image of comparable size, $64 \times 64$ pixels, each channel performs a multiplication close to a Toeplitz matrix approximately of size $64^2\times64^2$.Although its coefficients are not individually addressable but they depend on the phasemask displayed on the SLM. Even with this relatively simple setup the amount of parasitic reflections and aberrations (mainly from the SLM) makes it more reasonable to adopt amatrix formalism. This dependency, in the context of aberrations and imperfect conjugation, is only fully encompassed using the transmission matrix approach. Overall, this is equivalent to $64^4\cdot 60 \cdot 100 \approx 10^{11}$ multiply-and-accumulate operations (MACs) per second, comparable to a standard graphics card from the 2000's \cite{NvidiaGeForce200}.

\subsection{The alignment method}
Our method relies on the diffraction of light by a sharp edge on the SLM. Between two zones of different voltage on the SLM, i.e. different refractive index, the sharp change in refractive indices diffracts some light, which thus appears dark in a conjugate plane. Therefore, although the SLM is a phase modulator, any phase pattern results in a modification of the collected intensity in a conjugate plane. For instance, on Fig. \ref{fig:Alignement2}.c targets are displayed on the SLM with the phase difference varying from 0 outside the target to $2\pi$ at the center of each target, making the edges clearly visible. By using a camera positioned near a conjugate plane of the µDisplay and SLM, it is possible to align the SLM and the µDisplay manually. However, the precision of a manual alignment is limited.
 This edge can then be detected using standard detection algorithms, such as ellipse fitting provided by the \textit{OpenCV} Python package \cite{opencv_library} for fine-tuning the alignment. To improve detection efficiency, a set of concentric circles with alternating retardance of 0 or $2\pi$ is displayed on the SLM instead of a single circle. The image of a dark disk is observed, it is easier to fit than a circle.\\

The corresponding code is available on GitHub [\url{https://github.com/TTimTT/SLM-alignment}] along with a video showing the calibration process for a $N_{mux}=8\times 8$ grid.
To ensure the two modulators are perfectly conjugated, one must proceed as follows:

\begin{enumerate}
    
    \item \textbf{SLM and µDisplay axial} positioning. Set the magnification of the $4f$ system to exactly match the ratio of pixel pitches. \\
    \begin{figure}[h!]
        \centering
        \includegraphics[width=0.85\columnwidth]{"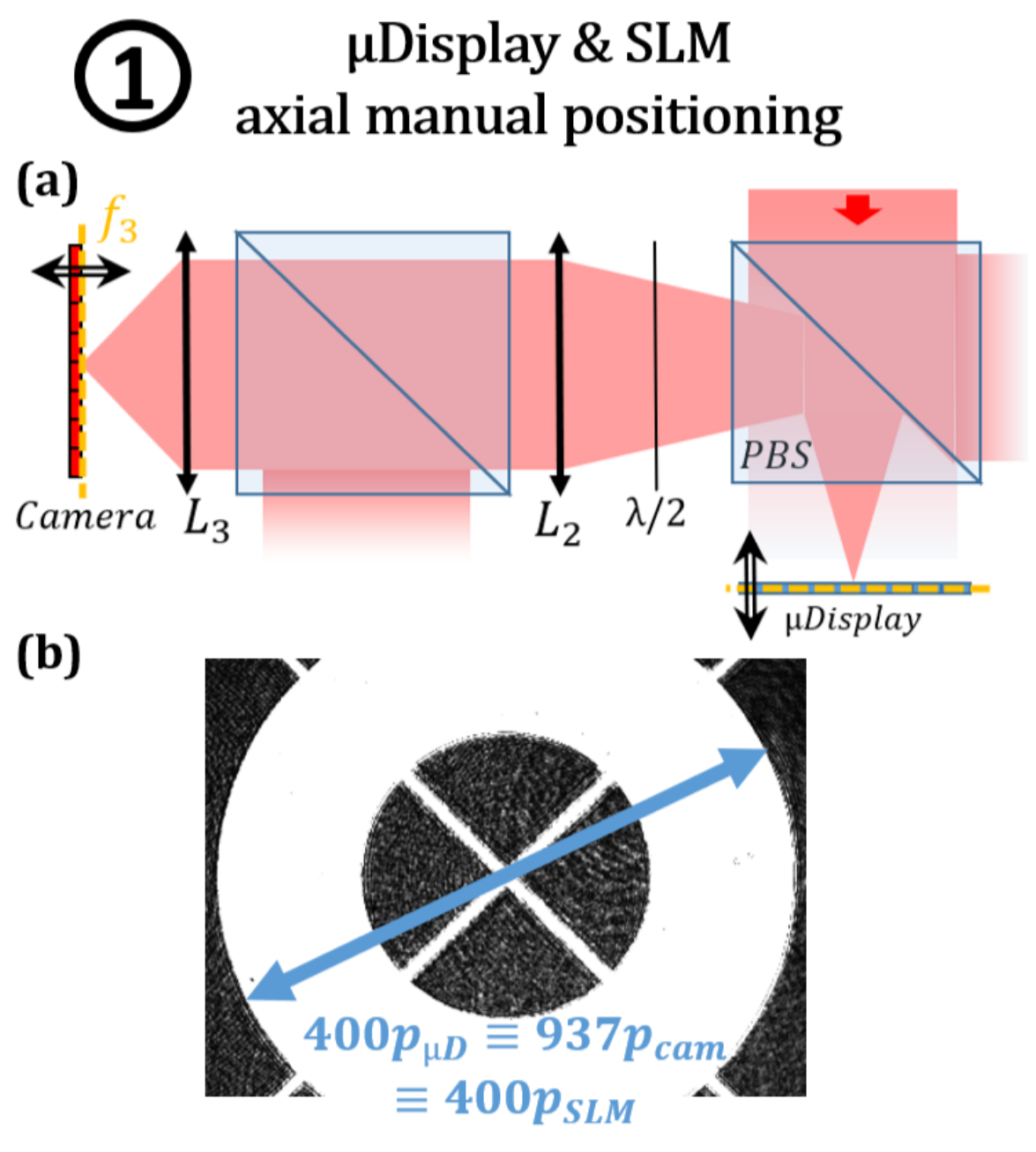"}
        \caption{\label{fig:Alignement1} \textbf{Adjustment of µDisplay and 4f-system position to obtain a given magnification }. \textbf{(a)} The SLM is replaced by a camera to locate the position of the µDisplay's image plane and adjust the magnification of the \textit{4f} system. \textbf{(b)} The magnification of the \textit{4f} system is set by displaying a circle of known width on the µDisplay and measuring the resulting image width on the camera.}
    \end{figure}
    As it is not possible to find two lenses with focal lengths in this exact ratio from standard catalogues, find the best possible combination to be close to the desired ratio. For this, adjust the relative positions of the two lenses and objects to build a $4f$ system with custom magnification. Position a camera at the conjugate plane of the µDisplay in place of the SLM (Fig. \ref{fig:Alignement1}.a). Adjust the distances in the \textit{4f} system $L_2L_3$ iteratively so that the observed image on the camera matches the required magnification (Fig. \ref{fig:Alignement1}.b). At the end of the process, measure the position of the camera's sensor relative to $L_3$. In our case, $f_2=100$mm and $f_3=200$mm. \newline
    Due to the two lenses not being distant of exactly $f_2+f_3$, the telecentricity of the system is broken. However, the corresponding field curvature is modest because the distance between the two lenses is in the vicinity of $f_2+f_3$: in our case the distance between the two lenses is $270$mm instead of $300$mm. Afterwards, we collect images from small neighboring zones of size $64^2$ pixels$^2$ on the µDisplay and SLM on the camera. The field from these zones thus differs by a spherical phase. Given the small size of these zones and the the low curvature of the wavefield, it can be approximated as a constant phase difference. In other words, neglecting any aberrations, these zones can be considered as \textit{isoplanetic patches}.

    \item \textbf{SLM manual alignment.}
    Roughly position the SLM at the previous location of the camera's sensor, close to conjugation with the µDisplay.
    \begin{enumerate}
        \begin{figure}[t!]
            \centering
            \includegraphics[width=0.7\columnwidth]{"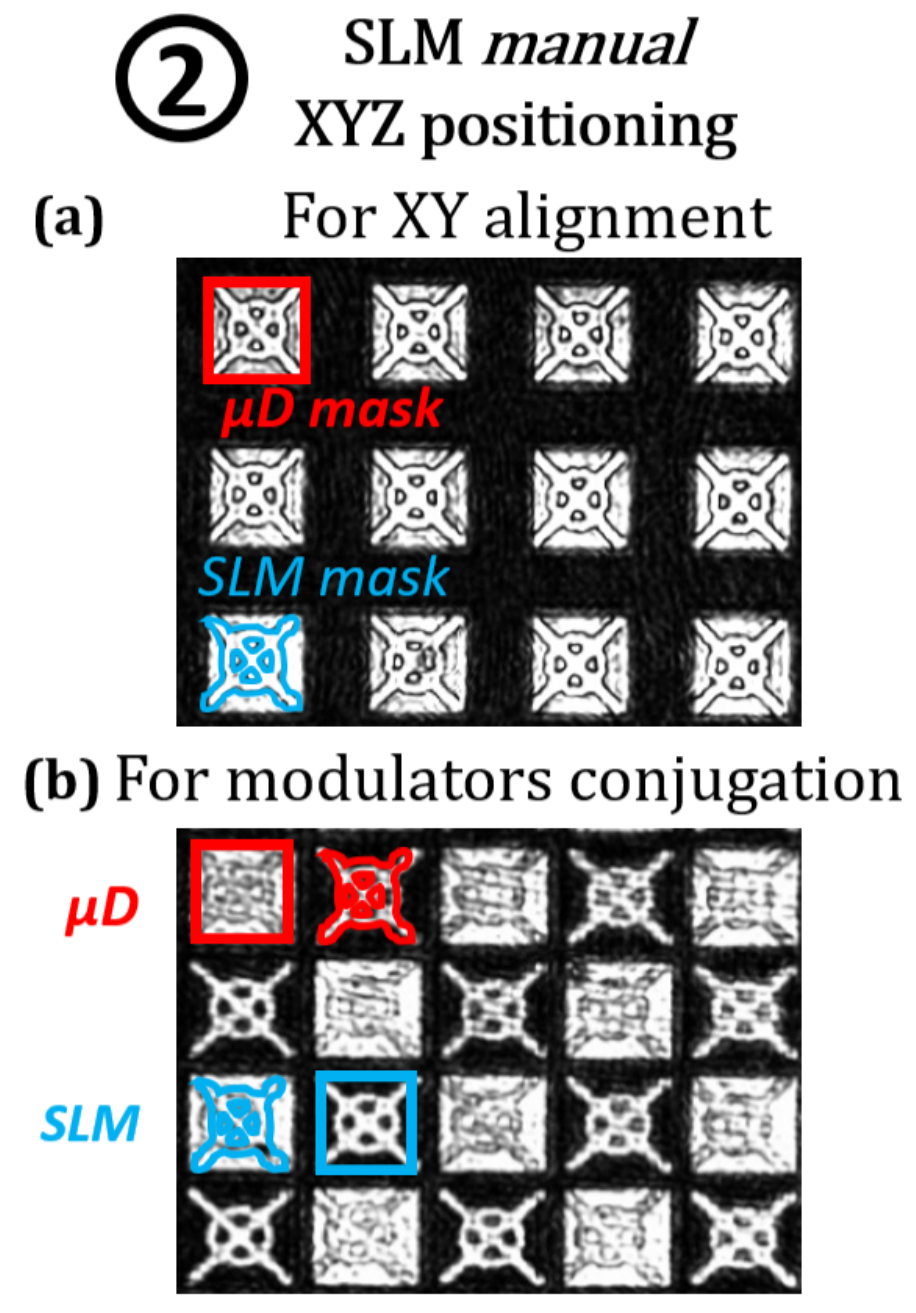"}
            \caption{\label{fig:Alignement2} \textbf{XYZ manual alignment using the superposition of complementary masks}. \textbf{(a)} Lateral positioning: targets displayed on the SLM are manually aligned with square of uniform intensities on the µDisplay. \textbf{(b)} Axial positioning: when the SLM is conjugated to the µDisplay, clear images of resolution targets are observed on both modulators.}
        \end{figure}
        \item Adjust the position of the SLM thanks to a six-axis translation and rotation stages (e.g. Thorlabs K6XS here) to align it  with the µDisplay. For this purpose, display a uniform square pattern on the µDisplay and targets on the SLM whose edges are visible due to the sharp change in refractive index (see Fig. \ref{fig:Alignement2}.a).
        \item Set the position of the SLM along the optical axis so that it is in the image plane of the µDisplay within the depth of field. Display alternate targets and squares of uniform intensity on the µDisplay and the opposite alternation on the SLM. Determine the position of the image planes of the SLM and µDisplay by translating the camera. Based on their relative positions, translate the SLM forward or backward along the optical axis (see SI). When sharp images of the targets from both the SLM and µDisplay are obtained (see Fig. \ref{fig:Alignement2}.b), it indicates the two modulators are nearly conjugated with a precision being the depth of field of lense $L3$.
        \item Iterate between the two previous settings.
        \end{enumerate}

    \item \textbf{SLM Automated alignments.} 
    \begin{enumerate}
        \begin{figure}[b!]
            \centering
            \includegraphics[width=0.8\columnwidth]{"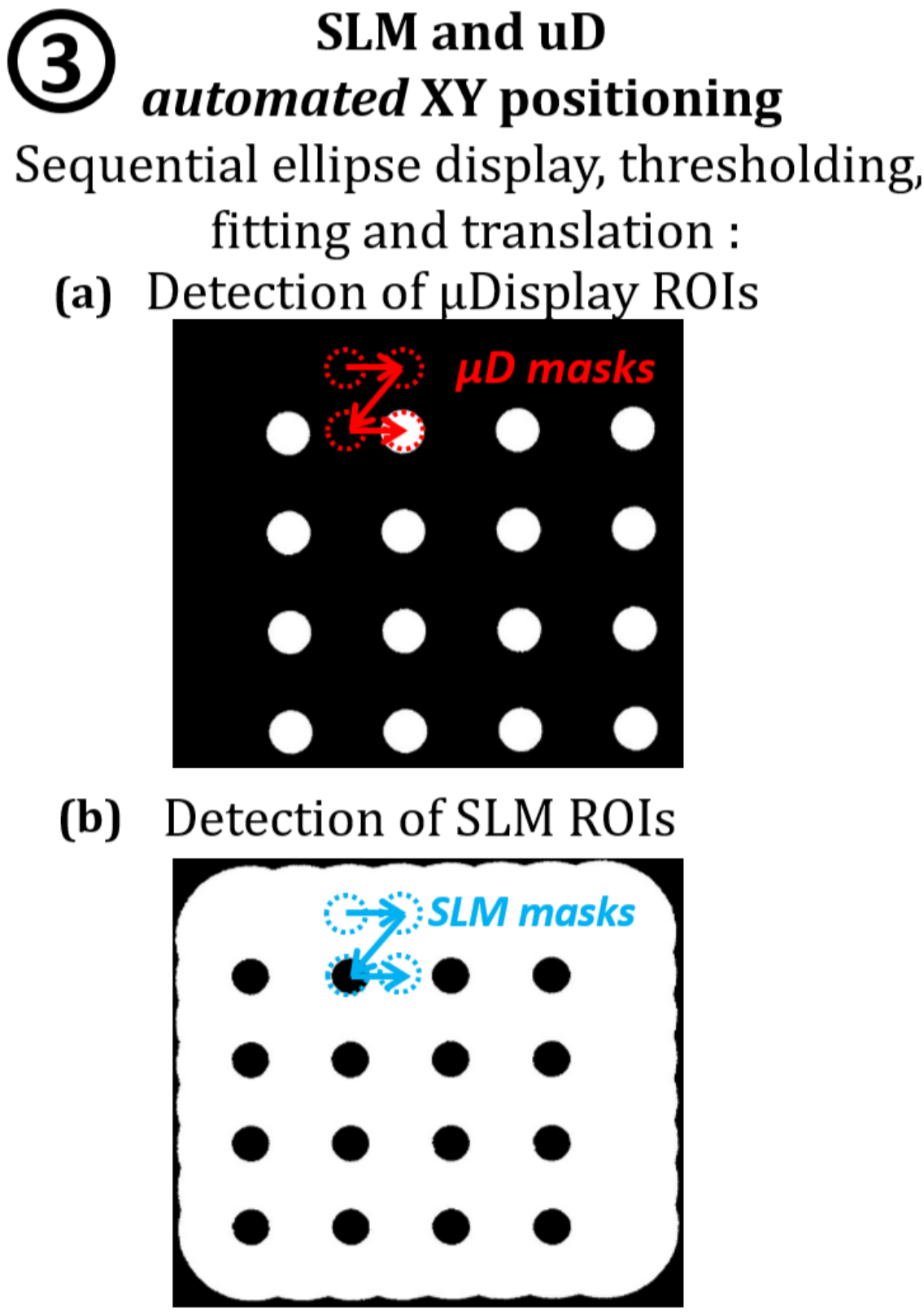"}
            \caption{\label{fig:Alignement3} \textbf{Automated XY fine positioning using sequential ellipse display, thresholding, fitting and translation : }\textbf{(a)} Disks of uniform intensity are displayed on the µDisplay. To avoid any overlap if the camera is defocused and minimize the flux, only one disk out of four is displayed at a time so that each disk's image never overlaps with their neighbors. Examples of the successive position of the disks for the four acquisitions are shown as red circles here. \textbf{(b)} Disks of alternating indices displayed on the SLM appears opaque on the camera due to wave diffusion by sharp edges and can be detected by standard ellipse fitting algorithm. Examples of the successive position of the disks for the four acquisitions are shown as blue circles here.}
        \end{figure}
        \item 
        \begin{itemize}
            \item Display disks on the amplitude modulator while maintaining a flat phase on the SLM (Fig. \ref{fig:Alignement3}.a).
        In the case of a defocused camera or an important misalignment between screens, only one disk out of four is displayed at a time, ensuring that each disk's image never overlaps with its neighbors and avoids any possible position detection mix up, resulting in four separate acquisitions.
        On Fig. \ref{fig:Alignement3}.a the successive positions of four detected disks are shown using dotted lines, illustration the acquisition scheme used. A video is available at [\url{https://github.com/TTimTT/SLM-alignment}].
            \item \label{ThresholdingSLM} Apply a binary thresholding operation. 
            \item Use an opening and closing morphological operation to remove the noise and smooth the disks.
            \item  Record the centres of the disks on the camera by performing ellipse fitting.
        \end{itemize}
        \item \begin{itemize}
         \item Display a uniform intensity on the µDisplay and display concentric patterns of alternate 0 and $\pi$ phase on the SLM.
        \item Apply thresholding and morphological operations again. The circle then appear as black circles on a white background. (Fig. \ref{fig:Alignement3}.b)
        \item Record the centres of the circles on the camera by performing ellipse fitting. 
        \end{itemize}
        \item Apply shifts by comparing the positions of the disks associated with the µDisplay and the SLM. 
        
        \item Repeat steps (a-c) until the algorithm converges.  

    \end{enumerate}
\end{enumerate}

    \begin{figure}[b!]
        \centering
        \includegraphics[width=0.6\columnwidth]{"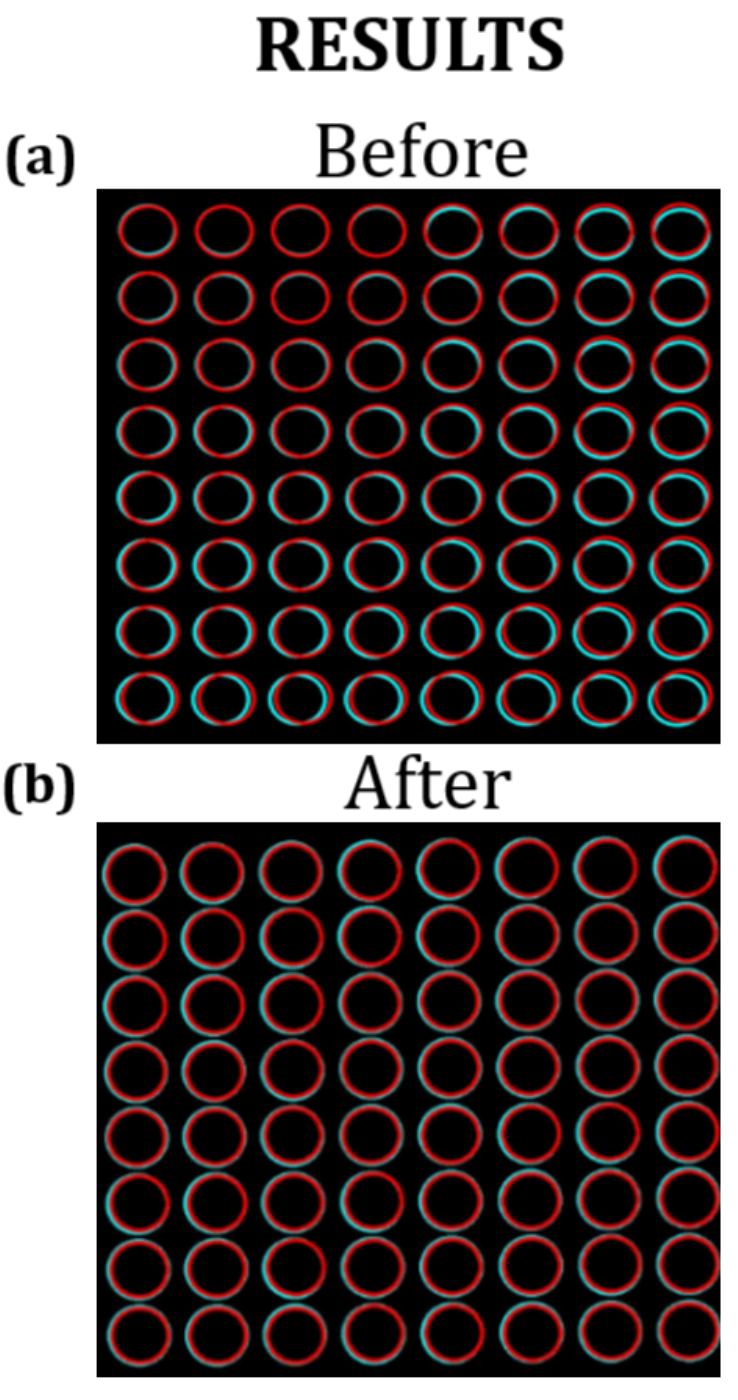"}
        \caption{\label{fig:AlignementResults} \textbf{Results of the calibration : }.\textbf{(a)} Initial calibrated surface: Each ROIs of the µDisplay (in red) and SLM could not be perfectly aligned manually due to limited precision in translation and rotation.\textbf{(b)} They match almost perfectly after using our algorithm.}
    \end{figure}

In the end, the RoIs of the µDisplay and SLM which were not matching (Figure \ref{fig:AlignementResults}.a) are shown on Figure \ref{fig:AlignementResults}.b and the center positions of each disk on the camera are recorded. To obtain the $(x,y)$ coordinates where the RoIs are located, the average disk width is calculated and extended by a few pixels (4 here) in each direction to account for any mismatch in the calibration, noise and defocus, giving the number of output pixels $N_{out} = N_{x,im}\cdot N_{y, im}$. For each acquisition, when a shot is taken, it is cropped at each recorded position. Finally, all the cropped images from one single shot are assembled as a batch of images with size $N_{mux}\cdot N_{x,im}\cdot N_{y, im}$. This format makes it readily usable by \textit{Pytorch} as it is designed for batch training. 

\section{Optical characterization}

\begin{figure}[b!] \centering
    \includegraphics[width=1.05\columnwidth]{"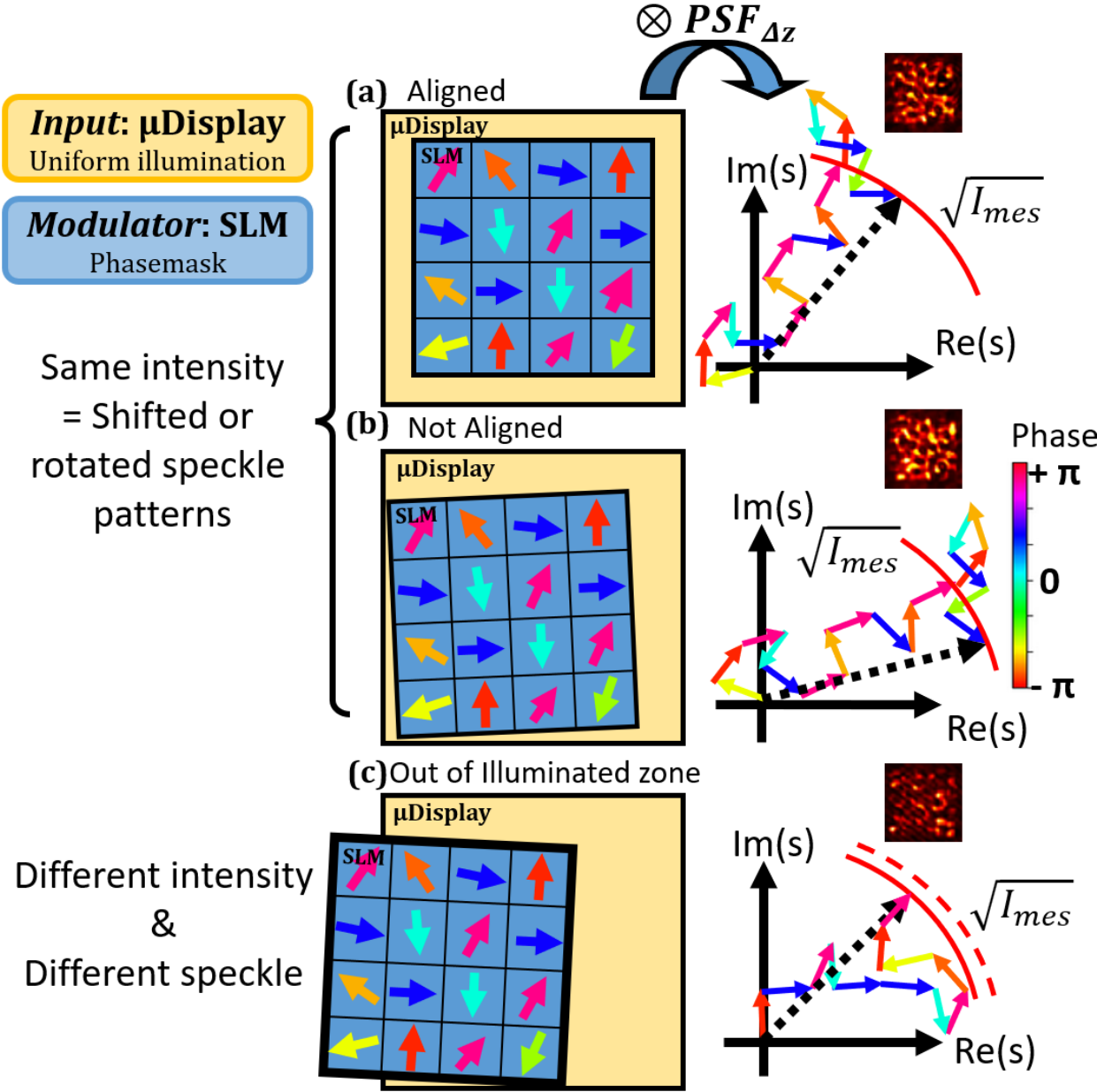"}
    \caption{\label{fig:IntensityMes} \textbf{Explanation of the intensity measurement alignment characterization} \textbf{(a)} After propagation of the field associated to a random mask and constant illumination to the defocused camera is a speckle pattern. The field at a given point is a sum of phasors. \textbf{(b)} For a moderate shift or rotation of the SLM relatively to the µDisplay, the same sum is obtained at another point thanks to the translation invariance. \textbf{(c)} For an extreme misalignment, the contribution of the phasors are modified and produces a different speckle pattern.}
\end{figure}

To assess whether this alignment procedure functions correctly and to quantify its precision, the accuracy of the ellipse detection is evaluated by acquiring the output images corresponding to a single illuminated pixel for each RoI over a wide range of defocus (see SI). This is equivalent to acquiring the modulus squared of the system's PSF which is also relevant for training a digital twin of this experiment \textit{in silico}. The precision of this algorithm is 1 pixel in the presence of noise and for various values of defocus (see SI). Next, the relative positions of the RoIs on the SLM with respect to those on the µDisplay are examined by applying a random phase pattern on the SLM and maintaining a constant illumination on the µDisplay, and analyzing the relative positions of the observed speckle-like pattern using autocorrelation. Autocorrelation is a well-known experimental method for tracking movement or deformation, employed for instance in Digital Image Correlation or Particle Image Velocimetry.

If a given RoI on the SLM is shifted or rotated relative to its corresponding RoI on the µDisplay compared to another set of µDisplay and SLM RoIs (i.e. \textit{a channel}) then the output speckle patterns of these two channels will be shifted or rotated relatively to the other. Indeed, at two different points, thanks to the circular symmetry of $PSF_{\Delta z}$ and thus the translation invariance of the configuration, a sum of phasor with the same amplitude is obtained up to a constant phaseshift due to the broken telecentricity (see Figure \ref{fig:IntensityMes}\textbf{a} \& \ref{fig:IntensityMes}\textbf{b}). This means that a different operation is performed on each channel. Otherwise, if the alignment is correct, the speckle patterns will have the same position, ensuring that all channels perform exactly the same operation. Hence, we can characterize this alignment using intensity measurements only. In the case of an extreme misalignment, the overlap between the µDisplay illumination and the SLM phasemasks do not match at all and instead of shifted identical speckle patterns, different speckle patterns are observed because the sum of phasor is modified (see Figure \ref{fig:IntensityMes}.\textbf{c}.)

As an experiment, identical random phase masks are displayed on the SLM for each RoI, with constant illumination from the µDisplay.

\begin{figure}[h!] \centering
\includegraphics[angle=270,width=1.02\columnwidth]{"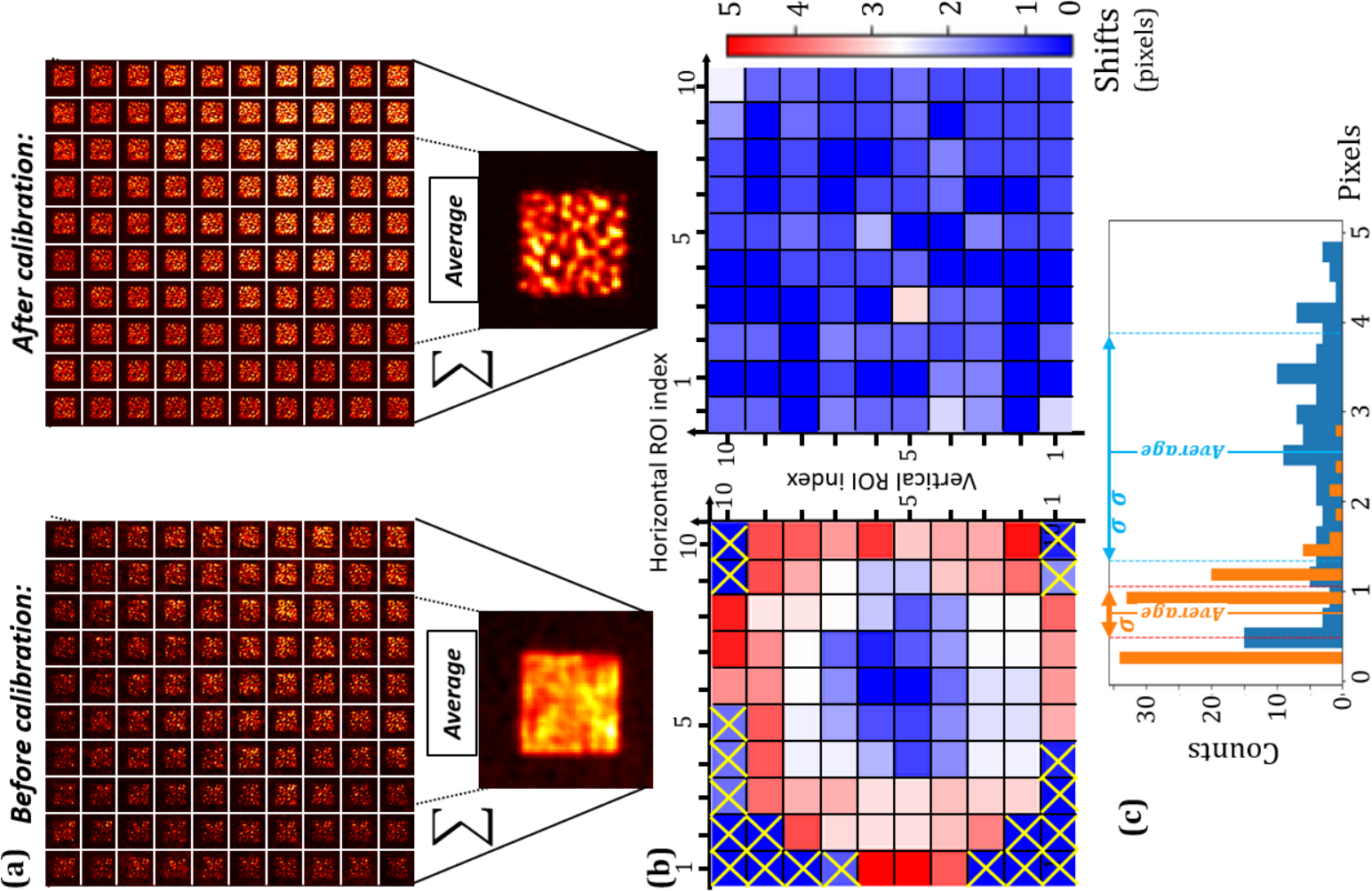"}
\caption{\label{fig:SpeckleAligned} \textbf{Comparison of the position of the speckle-like outputs before and after the automated alignment process.} \textbf{(a)} An example of raw speckle patterns and the average image. After calibration, the average image is a speckle-like pattern (right) whereas before calibration the sum averages to a square corresponding to the illuminated zone (left). \textbf{(b)} The number of shifts of each RoI compared to the central image measured with cross-correlation. The yellow cross indicates where the autocorrelation becomes irrelevant because the speckles patterns are too different. \textbf{(c)} A typical histogram of shifts: before the calibration, the images are shifted by up to 5 pixels (in blue). After calibration the images are shifted by less than 3 pixels, with the vast majority being shifted by 0 or 1 pixel (in orange).}
\end{figure}

 The intensity observed on the camera is shown on Fig. \ref{fig:SpeckleAligned}.a. The positions of the resulting output speckle patterns before and after the calibration process are compared by finding the maximum of the cross-correlation between each image and the central image at coordinate (6,6) on the $10 \times 10$ grid. Before calibration, the further an image is from the center, the more shifted each speckle pattern is (Fig. \ref{fig:SpeckleAligned}.b). Due to the initial misalignment, images too far from the center are too different for the cross-correlation to yield meaningful results, they are marked by a cross on Figure \ref{fig:SpeckleAligned}.b. As explained above using Figure \ref{fig:IntensityMes}, in this case the misalignment is too strong to obtain the same speckle pattern. This indeed happens for the farthest ROIs from the center as expected (Figure \ref{fig:SpeckleAligned}.b left). In this case, the maximum correlation between different speckles yields only the relative position of the illumination, which is 1 pixel here, determined by the precision of the ellipse fitting algorithm detailed in the SI. \newline 
However, after calibration, the images are shifted by less than 2 pixels, as shown in Fig. \ref{fig:SpeckleAligned}.b. This precision can be attributed to the precision of the ellipse fitting algorithm. After calibration, the speckle patterns are on average 0.8 pixels apart, compared to 2.5 pixels before calibration. As shown on the histogram \ref{fig:SpeckleAligned}.c, more than 90\% of them are below 1.5 pixels, which is below the diffraction limit, resulting in similar speckle-like outputs. Consequently, after calibration, the sum of each speckle-like pattern does not average out as a spatially incoherent sum, as seen in Figure \ref{fig:SpeckleAligned}.a., but instead forms a speckle-like pattern similar to that observed in each RoI. This means that each RoI is indeed usable as a parallel channel to process information optically.

 \section{Convergence of the algorithm}

\begin{figure}[b!] \centering
\includegraphics[width=1.05\columnwidth]{"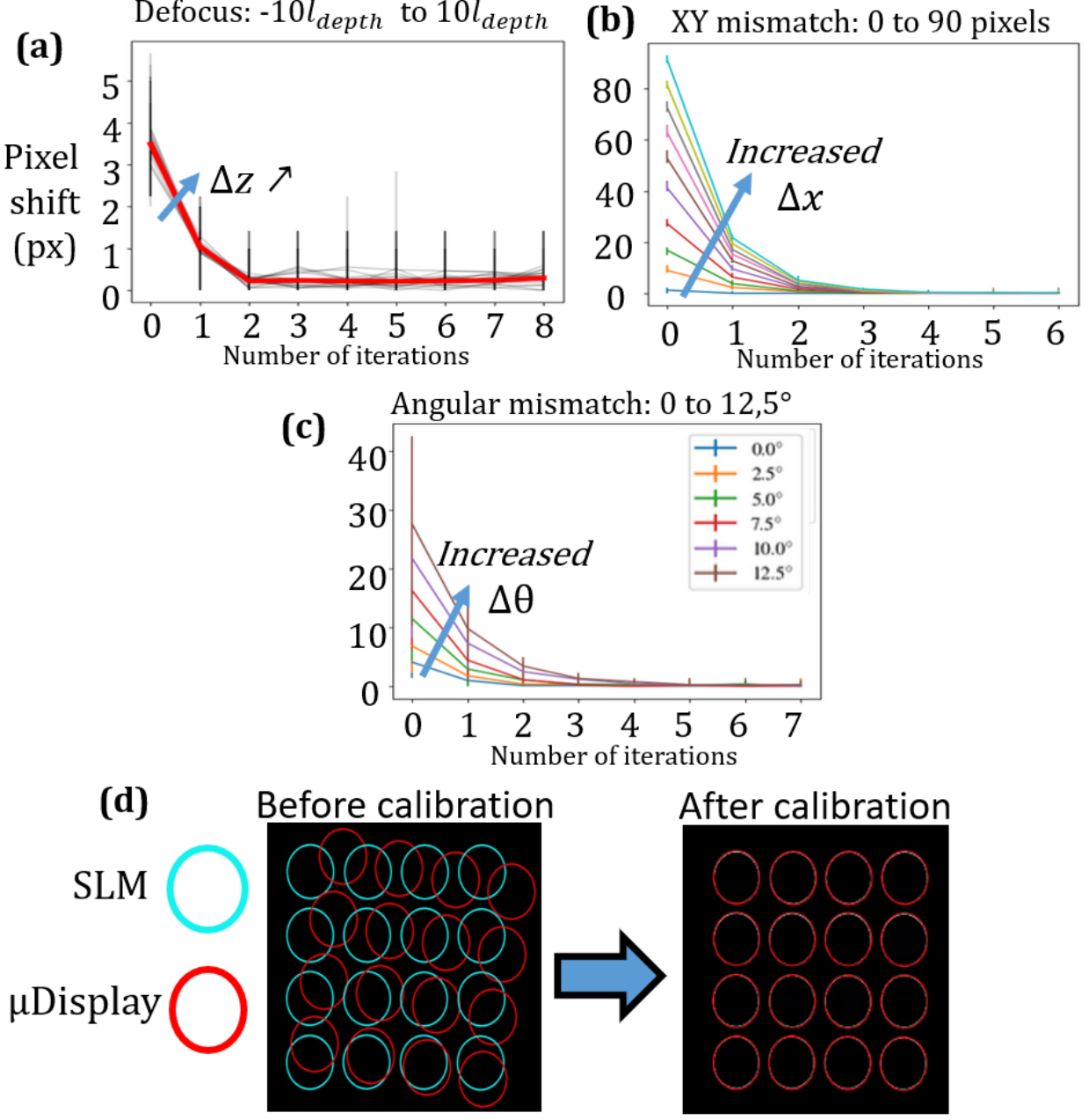"}
\caption{\label{fig:MismatchAlignment} \textbf{Evolution of the alignment with the number of iteration} \textbf{(a)} For a defocus increased gradually within $\pm$10 times the depth of field showing no influence on this parameter, each realization is shown in grey and the mean in red, the error bars show the variability between channels. \textbf{(b)} For increasing position mismatch. \textbf{(c)} For increasing angular mismatch \textbf{(d)} Results of the alignment procedure for a strong shift and angular mismatch for a $4\times 4$ channels configuration with $128\times 128$ pixels$^2$ ROIs.}
\end{figure}

In this section, we examine the convergence of the algorithm in a practical use case where the manual alignment as been achieved as best as possible. Then, the robustness of the algorithm to strong defocus, rotation and translation mismatch is examined. We suppose those degrees of freedom are linearly independent, each of them is manually increased gradually and the difference in the position of the detected ellipse of the µDisplay and SLM, the \textit{pixel shift}, is plotted on Figure \ref{fig:MismatchAlignment} \textbf{a} to \textbf{c}. Each curve corresponds to a given position of a translation or rotation stage and is an average over all ROIs ($4\times 4$ here), the error bars capture the variability among the channels for each experiment and the blue arrow shows in which direction the curve moves for an increase in the parameter.

First, we see that, as expected, our algorithm is not sensitive to defocus as seen on Figure \ref{fig:MismatchAlignment}.\textbf{a}, the convergence curve for each position is superposed over the mean for all positions (red curve). The convolution of a circular shape with a centrosymmetric PSF is indeed a circular shape, the detection of the dark spot's center is not affected by the amount of defocus used. The algorithm converges within 3 iterations and the error bars shows that the mismatch remains limited to $\sqrt{2}$ pixels given the precision of 1 pixels in each direction of the ellipse fitting algorithm. For shift and angular mismatch, longer convergence times are observed for stronger mismatch. However, in all cases, the algorithm converges within approximately 5 iterations corresponding to less than a minute and the error bars are barely visible proving that all channels are equivalently processed. Finally, we provide an example of convergence in a case of a strong shift and angle defocus on Figure \ref{fig:MismatchAlignment} \textbf{d}, it can be seen on the \textit{.gif} animation of the \href{https://github.com/TTimTT/SLM-alignment}{Github repository}. This video is live and shows that the convergence is obtained in less than a minute. 
However, the algorithm is sensible to the values of morphologies kernel and threshold used, particularly to the value of thresholding used at step \ref{ThresholdingSLM}. With the sharp-edge diffraction used here, the contrast of the SLM image over a bright background from the µDisplay uniform illumination is maximal. Lowering the contrast of the phasemask would even further increase the sensibility to this parameter. For this reason, we suppose it is not possible to use this algorithm with a deformable mirror.

\section{Advantages for a DNN}

\subsection{DNN structure and training}

The DNN examined consists of three layers, corresponding to three propagations through the setup. In ideal conditions, the output $y_n$ of the optical setup at optical layer $n$ can be written as a function of the input $x_n$, the phasemask $\phi_n(\overrightarrow{r},t)$ and the defocus kernel $PSF_{\Delta z}$:
\begin{equation}
    y_n(\overrightarrow{r},t) = | x_n(\overrightarrow{r},t) e^{j\cdot \phi_n(\overrightarrow{r},t)} \otimes PSF_{\Delta z}|^2
\end{equation}
In realistic conditions, i.e. with aberrations or parasitic reflections (see SI), it can simply be modeled as a multiplication with a transfer matrix containing the physical weights $W_P$ i.e.:
\begin{equation}
    y_n(\overrightarrow{r},t) = | W_P \cdot x_n|^2
\end{equation}
To train this DNN, we attempted the following:
\begin{itemize}
\item  Training a digital twin based on a physical model by fitting the experimental PSF (see SI) and transferring the weight updates to the experiment \cite{PNNbackprop_WrightMcMahon2022}.
\item  Training a digital twin using an experimental measurement of the transmission matrix in the Hadamard basis \cite{TM_PopoffGiganPRL2010}, with the loss incorporating the experimental output and transferring the weight updates to the experiment (see SI). Interestingly, thanks to our multiplexing method, it is possible to acquire the transmission matrices for $N_{ROI}=100$ RoIs of size $N_{in}=64\times64$ pixels$^2$ to the camera (an output image being approximately $N_{out}=80\times 80$ pixels$^2$) as quickly as for a single one. This measurement is completed in approximately 20 minutes and corresponds to $N_{in}\cdot N_{out}\cdot N_{ROI}= 64^2 \cdot80^2 \cdot 100 \approx 2.6 \cdot 10^9$ measurements.
\end{itemize}
As is already known from the state of the art \cite{NNtraining_Momeni24}, these methods, which are implicitly based on backpropagation are challenging to operate due to low noise tolerance and do not work in this case. We suspect that this system, which relies on full-field optics exhibits many parasitic reflections leading to input-dependent interference that makes training not a straightforward task, as the computed gradient differs too much from the experimental one. 
Finally, a supervised version of the forward-forward algorithm \cite{MomeniFF_Science2023} is implemented. With this algorithm, the setup is used as an optical encoder \cite{PassiveDL_FeiGiganCaoArXiv} and the learning is performed with a small digital layer. As developped earlier, this algorithm is adapted to using the ability of wave-based devices to perform fast mixing of information \cite{MomeniFF_Science2023} and a precise knowledge of the relationship between output and input is not necessary as the learning is performed digitally.
At each layer, we train the digital matrix $D_n$ to maximize the difference between positively and negatively labeled data by minimizing the cosine similarity of the ouput of each layer included in the following loss function :
\begin{equation}
    L_n= log(1+e^{\theta cos_{sim}(D_n y_n^+, D_n y_n^-)})
\end{equation}
where $\theta$ is an adjustable learning parameter and $cos_{sim}(X, Y) = \frac{\braket{X | Y}}{||X||^2\cdot||Y||^2}$ measure the equivalent of an angle between two vectors here corresponding to two images of size $N_x\cdot N_y$.
Between each pass, fully-connected layers containing trainable weights are updated. The output data from all the channels are multiplied by the same digital matrix at each layer. For a simple classification problem such as MNIST, this supervised learning algorithm requires two forward passes during training and ten forward passes during the inference. It is therefore slower than standard backpropagation. The detailed structure and parameters of the DNN are presented in the last section of SI. Even though the training is achieved using those hidden electronic layers, their hardware implementation is plausible \cite{XieScalableCoherentProcessor_Science25} even if a full-field configuration using the transmission matrix remains challenging \cite{brunner2025roadmapneuromorphicphotonics}. The future challenge for this setup would be of course to incorporate fully optical learning.

\subsection{Effect on training and inference time.}

The training time and score with the multiplexing properly calibrated is compared to the case where we do not use our calibration procedure.
Although an algorithm involving more propagation through the setup is used, the training time over 10 epochs is approximately 1 hour for 3 layers. Based on a 2 epochs experiment, a similar experiment displaying one image at a time takes around 4 days; the training time indeed decreases linearly with the number of images displayed. This makes it possible to train a DNN with optics in the visible range within a duration that is convenient for laboratory experiments. In addition, we posit that it might make the optical setup less sensitive to thermal or mechanical fluctuations since the training time has a characteristic time closer to the typical mechanical and thermal relaxation timescales.
The score obtained is around 85\% for MNIST digits classification dataset. Without the spatial coherence retrieved using our algorithm the DNN cannot learn properly. The inference score remains stuck around 30\%. This score is still relatively modest but significant, it means that the data can propagate through the device without significant loss of information. The limit of the score might come from residual misalignment as well as from the Forward-Forward algorithm which has not been extensively studied inherent limitations of the FF algorithm which have been recently pointed out \cite{Chen2025}.

\subsection{Effect on the noise}

\begin{figure}[b!] \centering
\includegraphics[width=1.05\columnwidth]{"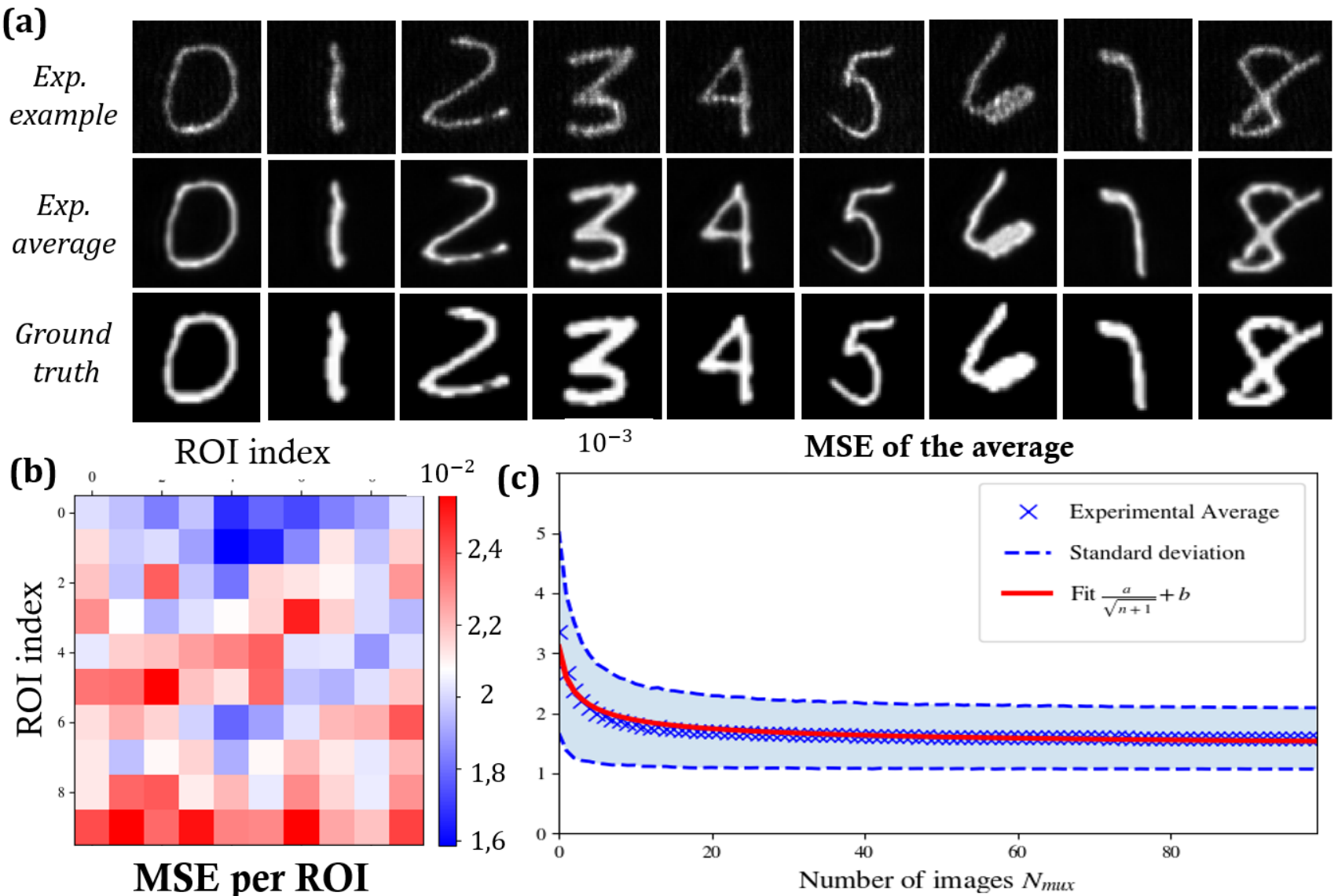"}
\caption{\label{fig:MSE} \textbf{Effect of the multiplexing on the noise} \textbf{(a)} Experimental acquisition of MNIST digits on the setup in a conjugated configuration. \textbf{(b)} MSE to the mean image for each RoI. \textbf{(c)} Evolution of the noise with the number of images taken into account for averaging with standard deviation and fit.}
\end{figure}

In addition to accelerating the training process of a DNN by two orders of magnitude, this multiplexing scheme also has the beneficial effect of reducing the noise level through averaging. A simple experiment to demonstrate this behaviour is to display figures from the standard MNIST database on the µDisplay while maintaining a flat phase on the SLM with the camera being conjugated to both the SLM and µDisplay (see Fig. \ref{fig:MSE}.a). This experiment is conducted over the entire MNIST digit dataset. First, the mean square error (MSE) for each channel, i.e. each RoI, is examined to verify whether some channels are more prone to noise. The MSE for each channel is computed and shown in Fig. \ref{fig:MSE}.b, indicating that each has a similar noise level. The evolution of the MSE with the number of RoIs used for averaging is plotted in Fig. \ref{fig:MSE}.c. It is fitted by a function of the form $a+\frac{b}{\sqrt{n}}$ with $a$ and $b$ as adjustable parameters. Good agreement is observed, demonstrating that part of the noise is indeed random. The noise decreases below a threshold of $2\cdot 10^{-3}$, reached for 30 approximately RoIs. This demonstrates another benefit of spatial multiplexing.

\subsection{Score by channel}

In this section, we examine if some channels are more successful than others at inference. If so, that would be a solid hint that some channels are performing different operations than their neighbors. To do this, we look at the inference score for each channels. All the scores range from 79\% to 86\% (see SI) proving that the channels are more or less equivalent. Yet a difference of 7\% is significant here: the score of 79\% being significantly lower than the state of the art while 86\% is an acceptable figure.

\section{Conclusion \& Perspectives}

In this article, we present a method to maximise the usable DoFs of an SLM to process large amounts of data in parallel leveraging the interesting scalability of full-field optics in terms of DoFs. Our setup consists of a conjugated amplitude and phase modulators receiving a plane wave, the output field is sent to a defocused camera. To achieve this, we have developed a procedure that ensures each channel effectively performs the same physical operation. This is verified using constant illumination and random phase masks. The presented procedure can be readily applied to the conjugation of additional spatial modulators by successively translating the position of each RoI on each modulator relative to the previously calibrated modulator. This approach can accelerate the training of diffractive neural networks based on spatial light modulators and reduce experimental noise, although the noise level remains high enough to hinder training with the backpropagation algorithm. It may also be useful in other active research fields in optics where the conjugation of several spatial light modulators is of interest. \newline
Training such a diffractive neural network is challenging especially if it relies on full-field optics and coherent light. It remains an open scientific question and is therefore a very active area of research \cite{NNtraining_Momeni24}. This experiment mimics propagation in layered media with non-linear effects and all the noise expected from such systems. It offers the scalability of full-field optics \cite{McMahon2023} in a highly flexible configuration, with data blending determined by the defocus level and the diaphragm diameter, the number of DoFs set numerically, high training speed, and a tunable noise level enabled by the spatial multiplexing described here. Therefore, we believe this is an ideal platform for investigating this scientific question and for bridging the fields of optics and machine learning in a quantitative manner. Experimentally adressing the limitations of spatial light modulators \cite{SLM_Jullien20} such as phase deformation from non-planarity \cite{Marco:20}, wavelength dependence\cite{BoldHaskell98}, pixel crosstalk\cite{Moser:19} but also parasitic reflections and multiple diffusion may provide a route to building efficient diffractive neural networks using wave propagation in structured media.

\begin{acknowledgments}
We thank Dr. Sébastien Popoff for his advises.
\end{acknowledgments}

\providecommand{\noopsort}[1]{}\providecommand{\singleletter}[1]{#1}%

\section*{Supplementary informations}

\subsection{Axial positioning of the SLM}

\begin{figure}[h!] \centering
\includegraphics[width=\columnwidth]{"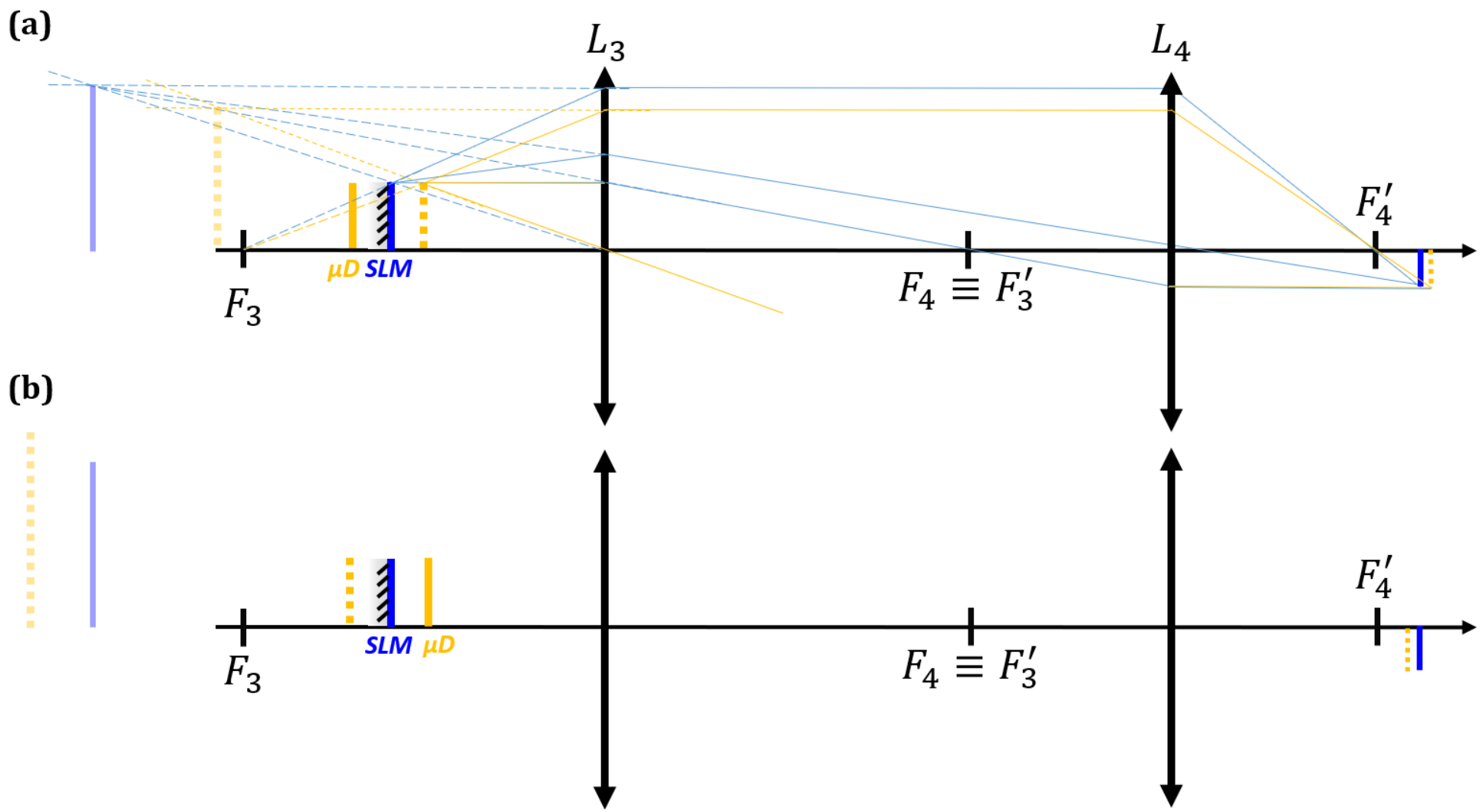"}
\caption{\label{fig:AxialPos} \textbf{Optical scheme for the \textit{4f} system conjugating the two modulators.} The µDisplay image from $L_1-L_2$ is indicated with a solid orange line at the left of the optical system, the SLM as a solid blue line with a mirror, the secondary image through the mirror and lense $L_3$ are represented with dotted lines.
\textbf{(a)} If the µDisplay image from \textit{4f} system $L_1-L_2$  is located farther from the lense than the SLM image, it means that the µDisplay image needs to be translated toward lense $L_3$. \textbf{(b)} If the µDisplay image from \textit{4f} system $L_1-L_2$  is located closer from the lense than the SLM image, it means that the µDisplay image (left, solid orange) needs to be translated away from lense $L_3$. }
\end{figure}

\begin{figure*}[t!] \centering
\includegraphics[width=0.9\textwidth]{"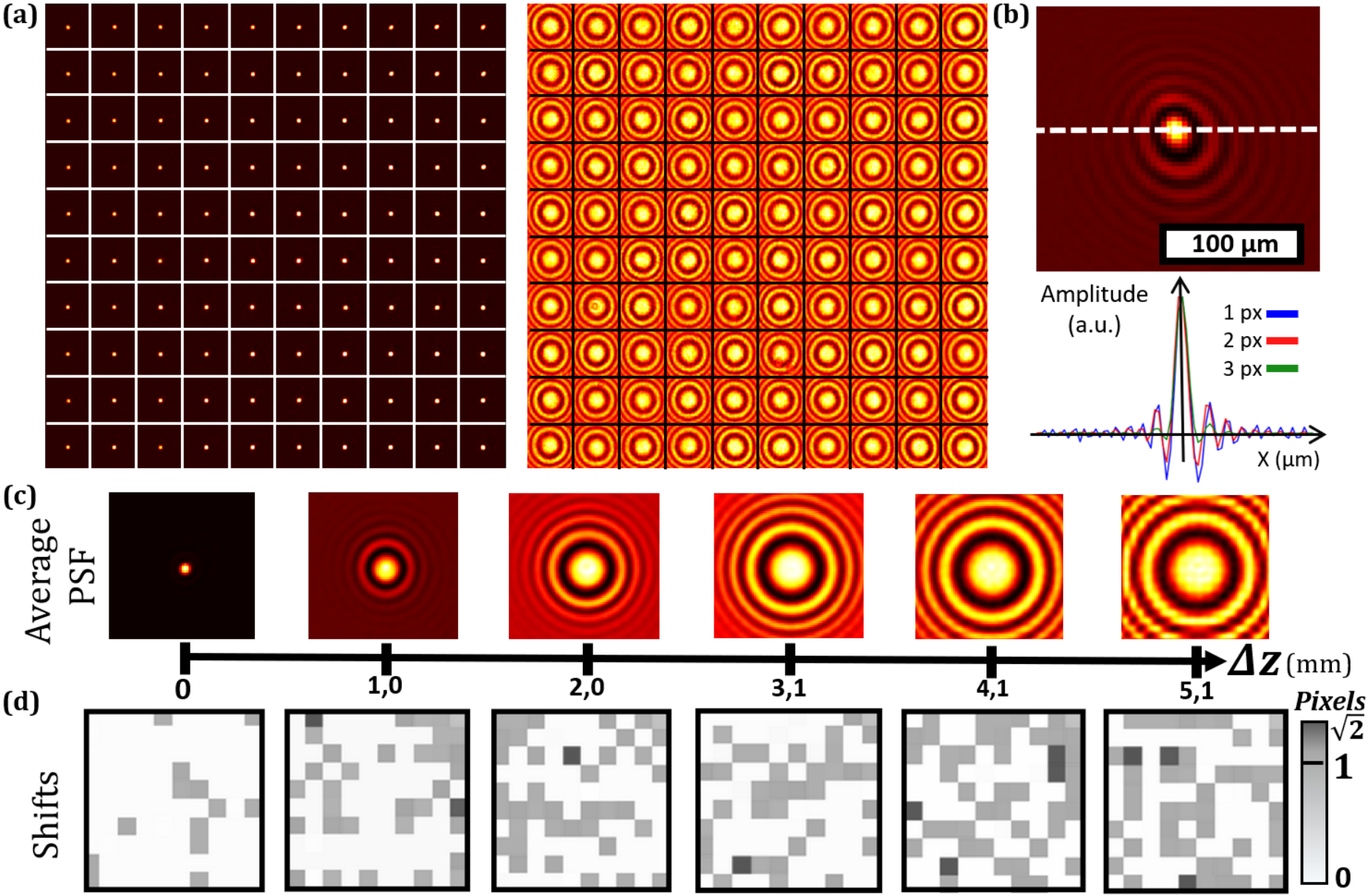"}
\caption{\label{fig:PSFacq} \textbf{PSF acquisition and alignment.} \textbf{(a)} PSF acquisition on each RoI for a conjugated (left) and defocused configuration (right). \textbf{(b)} The average PSF obtained by summing all the images on each RoI for a slight defocus. \textbf{(c)} The experimental PSFs averaged over all RoIs for a varying defocus. \textbf{(d)} Shifts for each RoI in pixels measured by autocorrelation.}
\end{figure*}

This step is not strictly necessary during the calibration procedure to function as it is robust to defocus, as shown previously. However, for the future of this experiment, in order to use machine learning and phase reconstruction algorithms it is useful to have a good initial guess of the field in this plane. Therefore, we aim to use the \textit{ simplest} possible configuration, in which the two modulators are approximately conjugated with a pixel-to-pixel matching. \\
Depending on the relative position of the images of the SLM and µDisplay after lens $L_4$, the relative position of the SLM and the µDisplay can be deduced, as shown in Fig. \ref{fig:AxialPos}. The reflection from the reflective layer of the SLM must be taken into account. The SLM is translated according to the observed relative position of the images along the optical axis. At each step, the camera must also be translated again to locate the relative position of the two images. This iterative procedure converges to a state in which the images from the µDisplay and the edges of the images displayed on the SLM appear sharp.

\subsection{Ellipse detection efficiency \& PSF acquisition}

Squares of increasing width are displayed on the µDisplay for various values of defocus. The SLM phasemask remains flat throughout the whole process. Similar PSF images are obtained on each RoI (see Fig. \ref{fig:PSFacq}.a) confirming the correct alignment.
For a single illuminated pixel, the PSF signal is above the noise but of comparable amplitude; for two pixels it is well resolved. From interference with an unmodulated background, the field is obtained as shown in the cross-section view (Fig. \ref{fig:PSFacq}.b), this is highlighted by the fact that signal decrease below the background value similarly to an homodyne interferometric scheme \cite{Marco:20}. For the conjugated case, the well-known Airy spot \cite{JMertz} is observed. For one and two pixels wide inputs, the two profiles are almost identical based on the positions of the zeros, as can be seen in the cross-section plot in Fig. \ref{fig:PSFacq}.b (only the levels of the first secondary lobes differ slightly). However for a three pixel wide input, the profile begins to enlarge. This indicates that three pixels is approximately the diffraction limit of this setup from the µDisplay plane and that it is possible to use $2\times 2$ macropixels while remaining diffraction limited in order to achieve a higher signal.   
Negligible aberrations are visible, they are slightly more important in the RoIs far from the central zone of the camera where the paraxial approximation is less valid.
For each value of defocus and input width, all images are summed to obtain average PSFs, as shown on Fig. \ref{fig:PSFacq}.c. As the noises differs for each image, it cancels out when averaging and the signal-to-noise ratio (SNR) increases roughly by a factor of 10 when computing the average PSF. 
For each value of defocus, starting from a perfectly conjugated configuration to a large defocus, where the PSF main lobe is approximately 30 pixels wide, the PSF is experimentally measured.
The position of the PSF obtained for each RoI relatively to the average PSF is then examined by autocorrelation for each value of defocus.  On Fig. \ref{fig:PSFacq}.d. is represented the measured shift for each RoI to the average PSF in pixels.  The maximum shift is 1 pixel in each direction leading to a maximum shift of $\sqrt{2}$ pixels. In any case this is lower than the diffraction limit. We assume, it is the limit precision of the autocorrelation calculation in the presence of noise as shown in the next section. The number of shifted images augments with the defocus but stays around 1 pixel.

The theoretical PSF can be computed by propagating the wave from the image plane, where the field is the Airy spot, to the camera plane over a distance $\Delta z$ using the Huygens-Fresnel principle. The predicted PSF match the experimental results (see below) 


\subsection{Analytical formulation of the PSF and PSF fitting}

The field from the image plane of the µDisplay is an Airy spot (whose dimensions are given by the limiting $N\!A$, here from lense $L_3$ with the largest focal length). Applying the Huygens-Fresnel principle, it is integrated to obtain the field at the observation plane at a distance $\Delta z$ [Born\&Wolf] :
\begin{equation} \label{eq: HF}
    PSF_{\Delta z}(\overrightarrow{r}) \propto \Delta z \cdot \iint PSF_0(\overrightarrow{r_0}) \cdot e^{j \frac{2 \pi}{\lambda} l(\overrightarrow{r_0})}\frac{dr_0}{l(\overrightarrow{r_0})^2}
\end{equation}
with $l$ being the distance between the observation point and integration point: $\overrightarrow{l} = \overrightarrow{r}+ \overrightarrow{\Delta z} -\overrightarrow{r_0}$ hence $l=\sqrt{((x-x_0)^2+(y-y_0)^2+\Delta z^2)}$ and:
\begin{equation}
PSF_0(\overrightarrow{r_0})\propto\frac{J_1(\frac{2\pi}{N\!A \lambda }r_0)}{r_0}
\end{equation}
from the integration of the field originating from a lense available in classical textbooks of imaging optics [JMertz]. Thus Equation \ref{eq: HF} writes:
\begin{equation} \label{eq:IntJ1}
\begin{split}
&PSF_{\Delta z}(\overrightarrow{r}) \propto \\ 
&\Delta z \cdot \iint \frac{J_1(\frac{2\pi}{N\!A \lambda }r_0)e^{j \frac{2 \pi}{\lambda} \sqrt{((x-x_0)^2+(y-y_0)^2+\Delta z^2)}}}{r_0\cdot((x-x_0)^2+(y-y_0)^2+\Delta z^2} dx_0dy_0
\end{split}
\end{equation}

\begin{figure}[h!] \centering
\includegraphics[width = 1.02\columnwidth]{"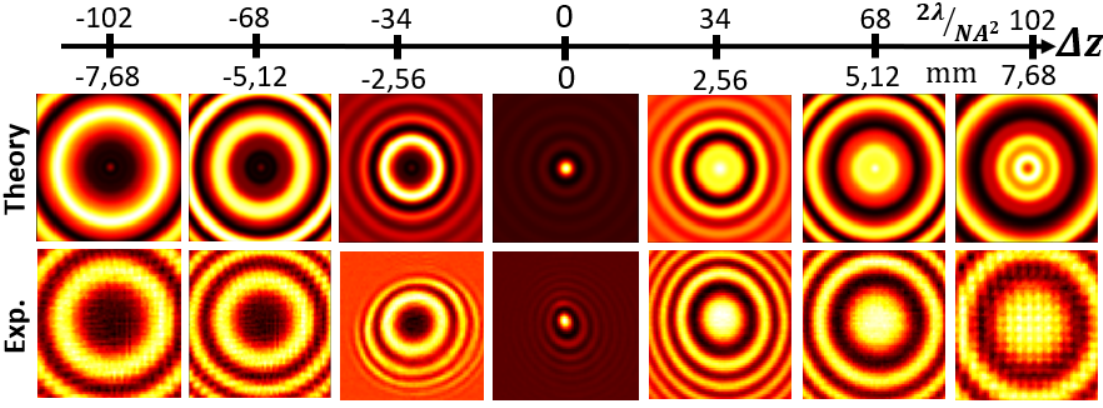"}
\caption{\label{fig:PSFfit} \textbf{PSF fitting.} the experimental PSF (top) compared to the semi-analytical predictions (bottom) for positive and negative defocus expressed in mm and in terms of the depth of field ($2\lambda/N\!A^2$).}
\end{figure}

On Fig. \ref{fig:PSFfit} are shown the experimental average PSF (top) from the measurement on each RoI (e.g. Fig. \ref{fig:PSFacq}.a.) and the theoretical PSF obtained from Equation \ref{eq:IntJ1} (bottom). A very good qualitative agreement is obtained between the two. 


\subsection{High frequency filtering and tunability of the speckle grain thanks to the diaphragm}

\begin{figure}[h!] \centering
\includegraphics[width=1.02\columnwidth]{"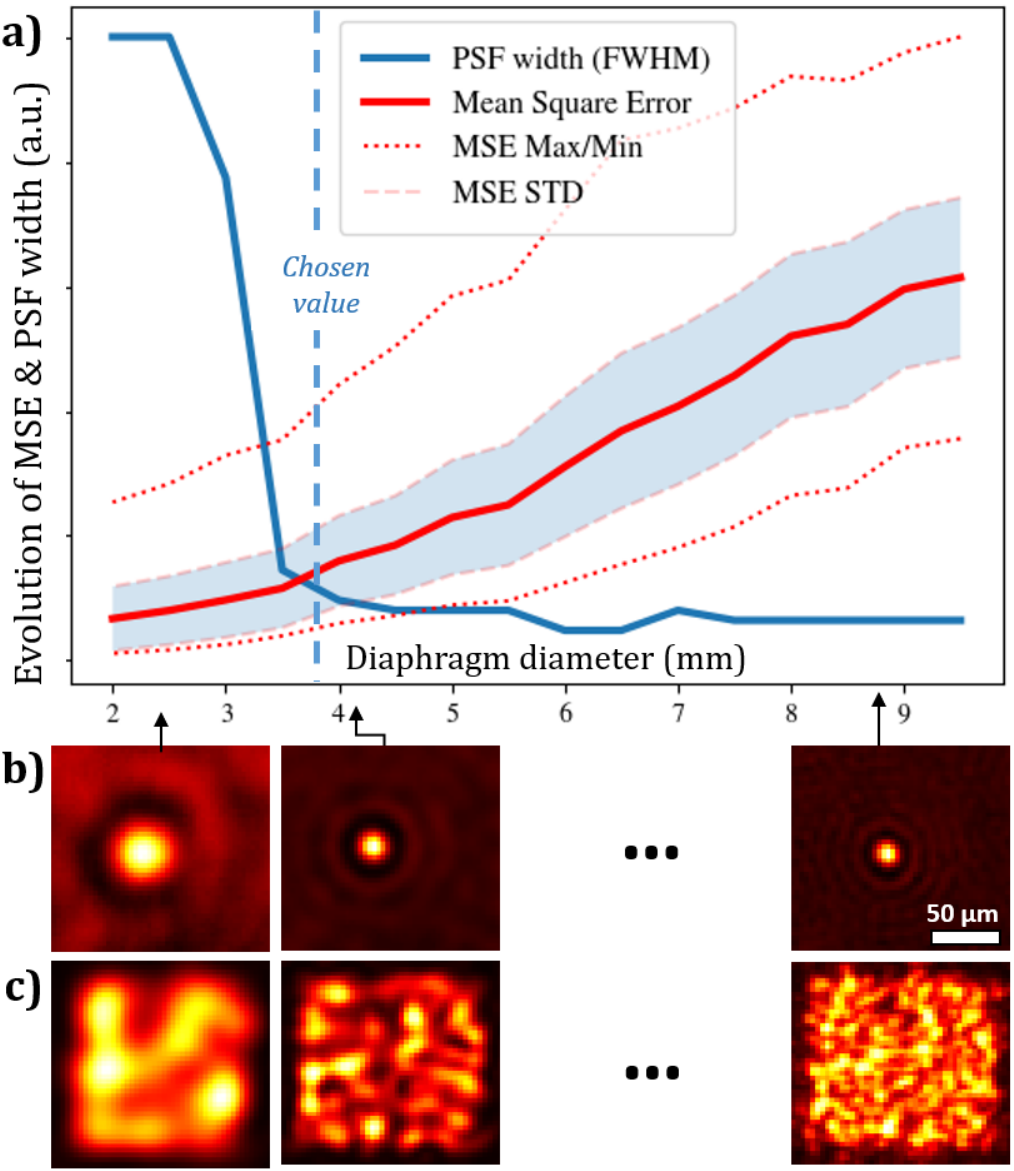"}
\caption{\label{fig:MSEDiaph} \small{\textbf{Effect of the diaphragm diameter:} Spatial variability of uniform inputs conjugated with random phase masks. \textbf{(a)} Spatial mean square error (MSE) and width of the point-spread function (PSF) as a function of the diaphragm filter $F_2$ diameter. \textbf{(b)} Experimental PSF acquisitions obtained by summing the PSF on all spatial channels. \textbf{(c)} Example of images with constant illumination on a square zone and random phase masks. A reduction of the diaphragm diameter increases the size of the speckle grain.}}
\end{figure}

Spatial light modulators are not perfect components. They exhibit spatial non-uniformities from fabrication, a filling factor below 100\% due to the necessary separation of liquid cristal pixels, and the voltage difference for each pixels is not perfectly confined in space to its area because of electromagnetic edge effects, leading to distorted wavefronts \cite{DMDsettingAberration}. The optical index difference between neighboring liquid-crystal cells introduces edges capable of diffracting light as a grating. This mainly contributes to high spatial frequencies content of the collected images and \textit{cross-talk} between RoIs as those index differences are sharp \cite{SLM_Jullien20, DynamicConfocal_Noetinger24}. This deviation from an ideal phase mask can be roughly modelled as a constant phase mask in the low spatial frequency range and a variable ,\textit{i.e.} phase mask input dependent, high frequency noise. To mitigate the effect of the low frequency mask, the RoIs are sized smaller than the characteristic length of the constant phase mask characteristic length. To filter these high frequencies and suppress cross-talk, a diaphragm $D$ is added in the Fourier plane of the modulators. We study the effects of this diaphragm diameter on the width of the PSF and the reduction of the MSE using constant illumination on the µDisplay and random phase masks on the SLM. The results are summarised up in Fig. \ref{fig:MSEDiaph}.a. Closing the diaphragm filters out high frequencies and spurious signals thus lowering the MSE without degrading the PSF until a certain threshold value is reached, at which point the PSF begins to enlarge (Fig. \ref{fig:MSEDiaph}.b). When this threshold is reached, it indicates that the NA of lens $L_6$ becomes the limiting NA, whereas previously it was lens $L_3$. Whether this enlargement is desirable remains an open question. On one hand, it results in a loss of high spatial frequency content from the input and this loss of information \cite{FourierO} might be detrimental to the classification. On the other hand, the convolution with a larger PSF - or, in machine-learning terminology, a larger kernel- blends the data more effectively leading to a coarser speckle grain (Fig. \ref{fig:MSEDiaph}.c). 

As a compromise, the diameter is set at $3.8$mm, just before this threshold. 


\subsection{Transmission matrix measurements}

\begin{figure*}[t!] \centering
\includegraphics[width=0.9\textwidth]{"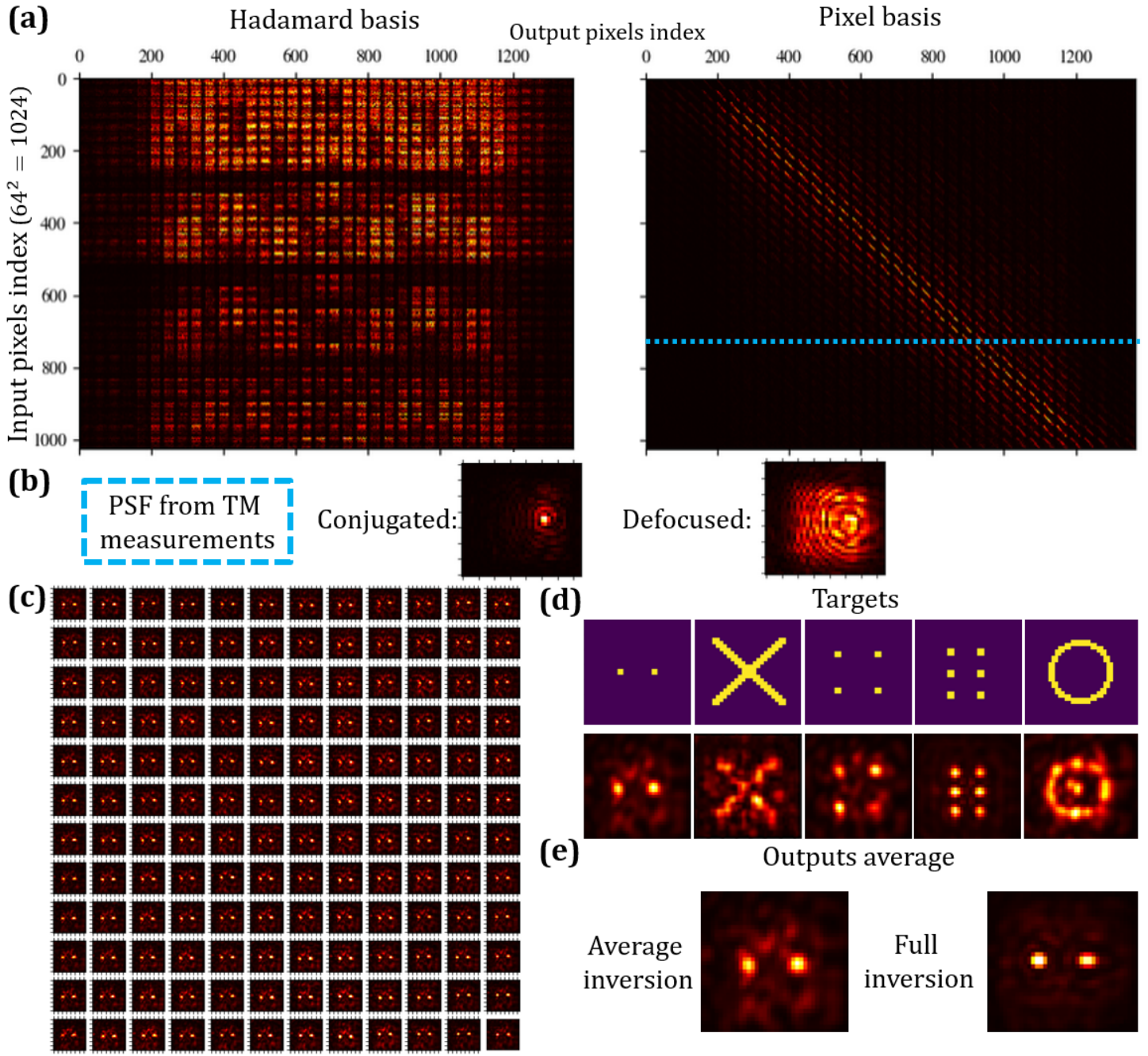"}
\caption{\label{fig:TMmes} \textbf{Transmission matrix measurements.} \textbf{(a)} The raw modulus of the transmission matrix in the Hadamard basis (left) and pixel basis (right). \textbf{(b)} One line of the TM represented as an image for conjugated (left) and defocused (right) configuration. \textbf{(c)} The refocusing on two points over all the RoI using the inverted transmission matrix averaged over all RoIs ($\pm =$ spatial average). \textbf{(d)} The output avergae over all RoIs of the refocusing on different patterns \textbf{(e)} Comparison of the average output obtained for a two point targets with inversion of the average matrix (right) or the inversion of all the transmission matrix obtained for each RoI (right).}
\end{figure*}

The transmission matrix from the SLM to the camera is measured over each RoI using the algorithm described in \cite{TM_PopoffGiganPRL2010}, with the reflection on the glass cover of the SLM serving as a reference beam. The measurement of the PSF, shown in Fig. \ref{fig:PSFacq}.b, indicates that this reflection is sufficiently strong to access the field by homodyne interferometry. As previously observed, the signal from $2\times 2$ pixels is similar to that from single pixels, but with a higher SNR in the PSF measurements. Therefore, we use the Hadamard basis on $64\times 64$ pixels RoIs with 2 pixels binning, meaning the basis contains $32\times32=1024$ elements. There are $10\times10$ RoIs so 100 transmission matrices are obtained. Theses matrices are similar one to another and can be condensed into an average matrix.
Using a change-of-basis matrix, the matrix is converted to the pixel basis as shown in Fig. \ref{fig:TMmes}.a. In the pixel basis, a matrix with a dominant diagonal is obtained, this is expected for a defocused optical system. For no defocus and moderate defocus, the lines of the matrix indeed corresponds to circular functions such as Airy spots, with or without defocus (Fig. \ref{fig:TMmes}.b).
To verify that the obtained matrix $M$ has physical significance, succesful refocusing on a given pattern is achieved across all RoIs, as shown on Fig. \ref{fig:TMmes}.c, d \& e. To achieve this, the matrix is filtered: only 10\% of the highest eigenvalues returned by the SVD algorithm are kept and a pseudo-inverse $iM$ is obtained by Tikhonov regularization $iM=(M^*\cdot M+SNR)^{-1}\cdot M^*$. The phase mask corresponding to a desired output is obtained by multiplying the output with $iM$. The refocusing on \textit{simple} sparse patterns such as two dots (Fig. \ref{fig:TMmes}.c \&d) is convincing, however when the patterns are more elaborate, such as a cross or a circle, the resulting output intensity exhibits significant deformations and artefacts (Fig. \ref{fig:TMmes}.d). This is explained by the fact that the flux equivalent to each pixel is too low to enable efficient focusing on complex patterns; as a result, the signal drops and the noise increases inversely \cite{Popoff2010}. This can be achieved using an average transmission matrix over all RoIs or for each RoI independently. In the latter case, the refocusing quality is significantly improved as can be seen in the average outputs depicted in Fig. \ref{fig:TMmes}.e, but the SVD, filtering, and inversion time increases from less than a minute to more than an hour, approximately scaling with the number of RoIs.\\

\begin{figure}[h!] \centering
\includegraphics[width=1.02\columnwidth]{"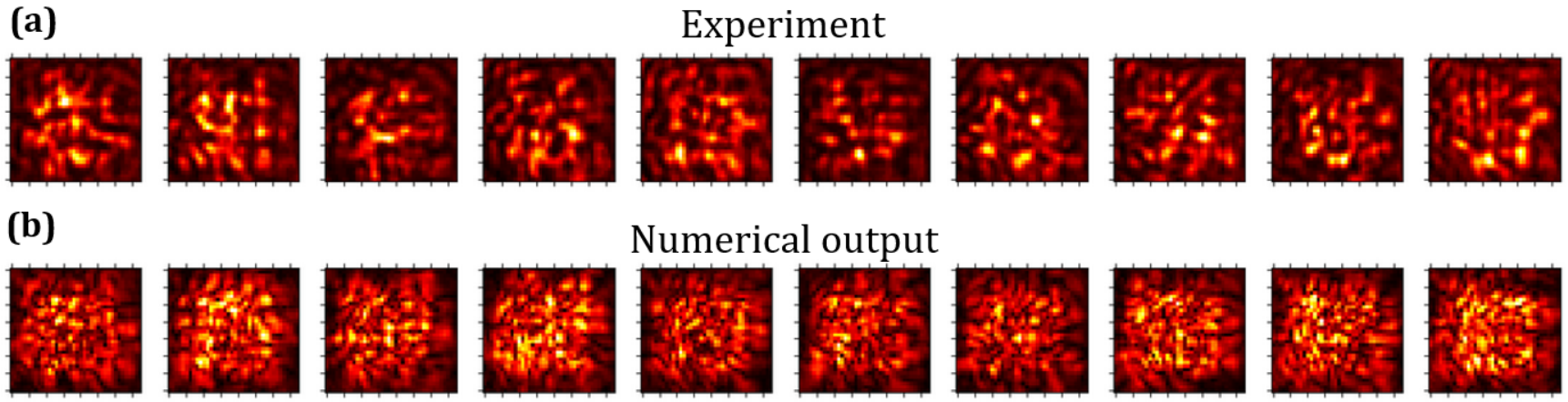"}
\caption{\label{fig:TMpred} \textbf{Random mask output} \textbf{(a)} Experimental. \textbf{(b)} Theoretical, from the measured transmission matrix output.}
\end{figure}

However, for random phasemasks displayed on the SLM, it is impossible to predict the output of the setup thanks to the transmission matrix. On Fig. \ref{fig:TMpred} experimental outputs for a random phasemask on the SLM are depicted along the predicted output using the measured transmission matrix. There is a strong discrepancy between the two. Any filtering using the SVD operator has no influence on this result.\\

Our guess is that this phenomenon, already known from the literature \cite{TM_PopoffGiganPRL2010, Popoff2010}, hinders the possibility to compute a relevant phasemask using direct backpropagation, either using a convolutive model (with an analytical formula using a fitted version of the measured PSF at the previous section) or a model using the experimental matrix. 

\subsection{Detailed neural network structure}

\begin{figure*}[t!]
\centering\includegraphics[width=\textwidth]{"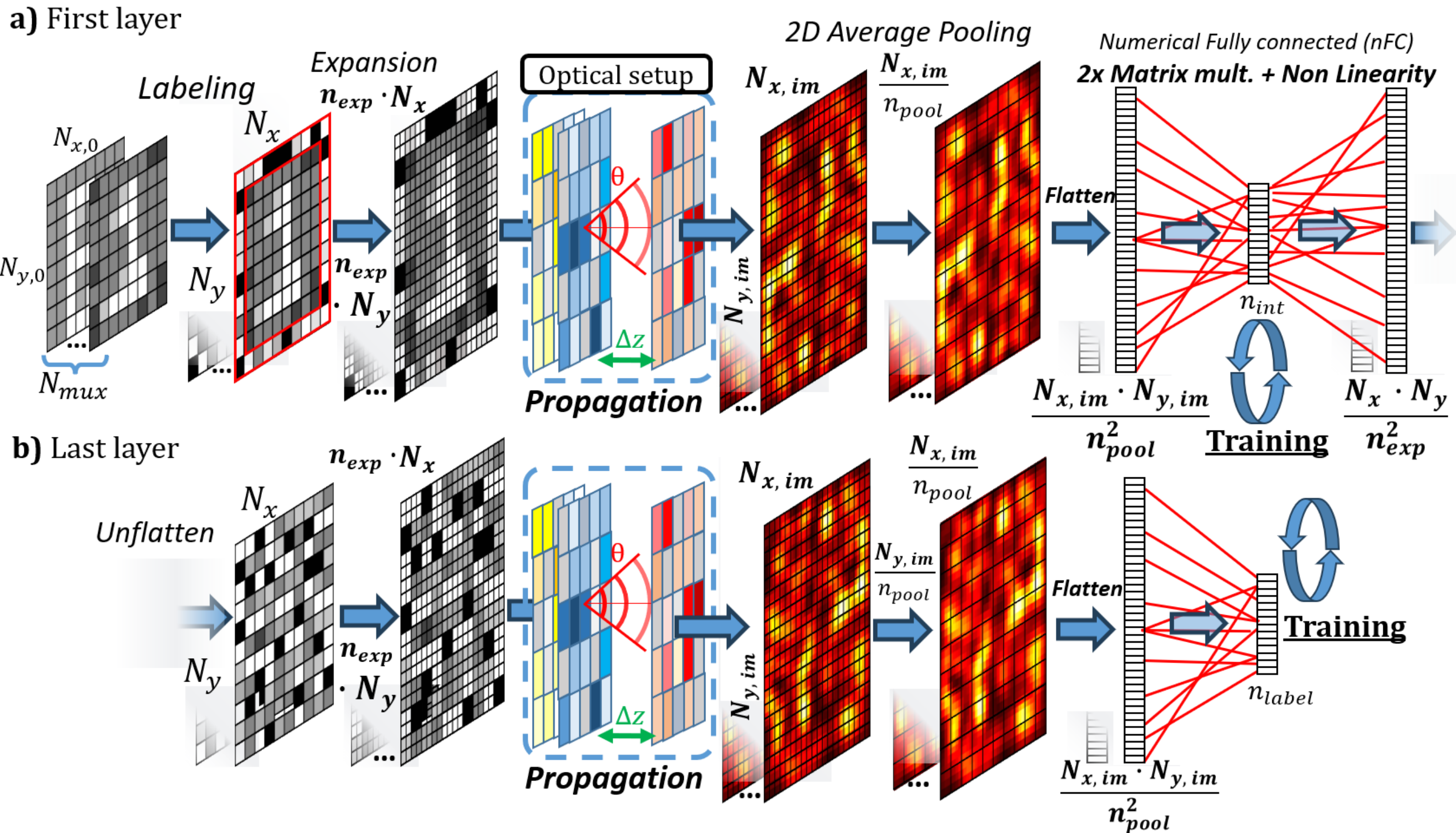"}
\caption{\label{fig:FullNN} \small{ \textbf{Full neural networks} depicted as a data pipeline for a depth $n_l=2$.} A batch of $N_{mux}$ images of size $N_{x,0}\times N_{y,0}$ is labeled and goes through the optical setup using the multiplexing procedure described in the main text. The resulting images are average pooled by a factor $n_{pool}$ to match the speckle grain size and sent to numerical fully connected layers before being sent again to the setup for additional optical propagation steps. At each step, the data is expanded by a factor $n_{exp}$ so that the information from one pixel is sent to more pixel to increase the flux and decrease the size of the intermediate numerical fully connected. The last layer maps the optical output to the number of expected categories $n_{label}$}
\end{figure*}

After trying training with backpropagation, we implemented a Forward-Forward algorithm \cite{MomeniFF_Science2023} where the optical setup is used as an optical encoder with fully connected layers in between two propagation on the optical setup. The full pipeline is represented on Fig. \ref{fig:FullNN}.

\paragraph{\textbf{Data labeling.}} The data has to be labeled for the Forward-Forward algorithm to recognize correctly \textit{vs.} incorrectly labeled data. To increase the inference score, we increase the energy associated to the label relatively to the energy associated to the data in terms of optical flux using\textit{ intrinsic class patterns} \cite{FF_Symba23}. It consists, for each label, in including the image in a random binary pattern instead of a one-hot vector. Therefore, the label information hidden in the data corresponds to the flux of many pixels instead of a single one.

\paragraph{\textbf{Phase masks.}} In this experiment, we use only static phase masks $\phi^{(l)}$ identical for all subimages on the SLM. An interesting question would be to determine experimentally the best mask to differentiate the positive from the negative data but it is beyond the scope of this paper.

\paragraph{\textbf{Numerical layers.}} The size of the transmission matrix in the numerical fully-connected (nFC) layers depends on the size of the subimages acquired $N_{pix, in}$ and the size of the output data to be sent again in the setups $N_{pix, out}$.
These two \textit{hyperparameters} can be optimized for each applications thanks to the flexibility of our setup. They could be used to perform data \textit{augmentation} or \textit{reduction} physically.
Although the objective of this paper is not to study this option in detail, we provide an example of such an optimization in the results section below.\newline
For the chosen value of the diaphragm $F_2$ diameter, the speckle grain size can be adjusted from 3 to 10 pixels using the diaphragm diameter.
With the chosen value the speckle grain size is approximately 5. To reduce the size of the nFC matrix, an average 2D pooling is added with a reduction factor $n_{pool}$ of 2 smaller than the speckle grain size ensuring that there is no loss of information.
Conversely, for the next propagation step, the output is expanded using the opposite operation of \textit{expanding} where one pixel is converted into $n_{exp}\cdot n_{exp}$ pixels.
Experimentally, this makes the flux associated to one output of the nFC more important compared to the noise level.
In addition, from the PSF measurements, we know that the setup is still diffraction limited for $2\times2$ pixels inputs and that $3 \times 3$ is an acceptable limit (see SI). $n_{expand}$ is therefore set to 2.

\paragraph{\textbf{Hyperparameters optimization.}} For different values of defocus, different values of hyperparameters may provide the best inference score.
Hence, to provide an accurate comparison of the physical parameters influence, numerical hyperparameters giving the best inference score such as the choice of loss and activation function and learning rates must be chosen for each physical configurations.
To automate this hyperparameter optimization, the \textit{Optuna} package \cite{Optuna_Akiba2019} is used.

\end{document}